# Evolution of Spots and Stripes in Cellular Automata

**Author:** Peter D. Turney, peter.turney@gmail.com

## Abstract

Cellular automata are computers, similar to Turing machines. The main difference is that Turing machines use a one-dimensional tape, whereas cellular automata use a two-dimensional grid. The best-known cellular automaton is the Game of Life, which is a universal computer. It belongs to a family of cellular automata with 262,144 members ($2^{18}$). Playing the Game of Life generally involves engineering; that is, assembling a device composed of various parts that are combined to achieve a specific intended result. Instead of *engineering* cellular automata, we propose *evolving* cellular automata. Evolution applies mutation and selection to a population of organisms. If a mutation increases the fitness of an organism, it may have many descendants, displacing the less fit organisms. Unlike engineering, evolution does not work towards an imagined goal. Evolution works towards increasing fitness, with no expectations about the specific form of the final result. Mutation, selection, and fitness yield structures that appear to be more organic and life-like than engineered structures. In our experiments, the patterns resulting from evolving cellular automata look much like the spots on leopards and the stripes on tigers.

## 1. Introduction

A large and dedicated community is devoted to engineering interesting structures in the Game of Life and in the many other life-like cellular automata. It is our hope that some of the readers of this article will look beyond *engineering* and explore what *evolution* can contribute to cellular automata.

Cellular automata began with the work of Stanislaw Ulam and John von Neumann in the 1940s (Poundstone, 2013). A cellular automaton is a regular grid of cells, typically square cells in two-dimensional space. Each cell can be in a finite number of different states, which are usually distinguished by colours or numbers. The grid is initialized by setting the starting states for each cell. The states of the cells change at integer intervals according to a given set of rules. Cells at time *t* simultaneously change their states at time *t* + 1. The new state for a given cell at time *t* + 1 is a function of the states of the cell's neighbours in the grid at time *t*.

The most popular cellular automaton is the *Game of Life*, introduced by John Horton Conway in 1970 (Gardner, 1970). At first, the game was played manually, on a Go board. When computers became common, the Go board was replaced with a computer screen, displaying a grid of black and white squares. At the beginning of a game, the squares are all white. The solo human player then creates a pattern by changing some of the white squares to black squares. The computer then animates the screen with a sequence of colour changes, from black to white and from white to black, following a set of deterministic rules. The objective of the game is to create an interesting sequence of images. As the game runs, black shapes grow and shrink. Some black shapes crawl across the screen, traveling in a cyclic sequence of motions.

The Game of Life appears to be simple, but it has been proven that the game is a universal computer (Berlekamp *et al.,* 1982). That is, anything that can be computed with a standard digital computer can be computed in the Game of Life. Enthusiasts have spent more than 50 years exploring the possibilities of the Game of Life. This exploration involves much trial and error, with a slow accumulation of expertise. There are some software tools that support this research, but the most interesting discoveries generally come from trial and error. Johnston and Greene (2022) provide an excellent introduction to the discoveries that Game of Life enthusiasts have made in the last half-century.

A difficulty we face with the Game of Life is that a small change to a Game of Life seed pattern can result in a large change in how the seed grows. In general, there is no easy way to direct the growing pattern to a desired outcome. This is where an evolutionary algorithm can be helpful. Consider that an acorn is quite different from a fully grown oak tree, yet evolution is able to create a path for growth that begins with an acorn and ends with an oak tree.

The general approach of evolutionary algorithms is to make small random changes to a structure and measure the impact of these changes. If a change improves the performance of the structure, the change is kept, although it might be dropped later. If a change reduces the performance of the structure, the change may be rejected, although it might be restored later.

Evolutionary algorithms can achieve results that are similar to the constructions made by Game of Life enthusiasts, yet evolutionary algorithms do not reason like humans. Just as biological evolution has created extremely complex and capable organisms without using planning or reasoning, evolutionary algorithms can create complex computational structures and patterns without planning and reasoning.

In Section 2, we introduce Conway's *Game of Life*. Section 3 explains how evolutionary algorithms can be used to evolve patterns in cellular automata. The aim is to reduce the human effort required to create interesting patterns, by allowing evolution to select organisms according to their fitness. Section 4 shows the results of our experiments with evolving cellular automata. Section 5 considers the relation between engineering and evolution. Section 6 discusses related work, Section 7 considers future work, and Section 8 summarizes the conclusions.

# 2. Conway's Game of Life

The Game of Life was originally played on a Go board with Go stones. Martin Gardner (1970) presented John Conway's *Game of Life* as three desiderata and three rules. The desiderata are (Gardner, 1970, p. 120):

1. There should be no initial pattern for which there is a simple proof that the population can grow without limit.
2. There should be initial patterns that *apparently* do grow without limit.
3. There should be simple initial patterns that grow and change for a considerable period of time before coming to end in three possible ways: fading away completely (from overcrowding or from becoming too sparse), settling into a stable configuration that remains unchanged thereafter, or entering an oscillating phase in which they repeat an endless cycle of two or more periods.

The three rules are as follows (Gardner, 1970, p. 120):

1. Survivals. Every counter (playing piece) with two or three neighboring counters survives for the next generation.
2. Deaths. Each counter with four or more neighbors dies (is removed) from overpopulation. Every counter with one neighbor or none dies from isolation.
3. Births. Each empty cell adjacent to exactly three neighbors—no more, no fewer—is a birth cell. A counter is placed on it at the next move.

Our terminology is different from Gardner's terminology. Gardner's *counter* corresponds to our *live black cell* and Gardner's *empty cell* corresponds to our *dead white cell*. This change in terminology comes from abandoning the Go board and switching to the computer screen.

The Game of Life belongs to a family of cellular automata with 262,144 members ($2^{18}$). There is a convenient way of representing the members of this family. The rule for Game of Life is compactly represented as B3/S23. B3 means a new cell is *born* (B) if exactly three of its eight neighbouring cells are alive. S23 means that an existing cell will *survive* (S) if either two or three of its eight neighbouring cells are alive.

In this article, we will explore three different cellular automata, B3/S23 (the Game of Life), B3678/S23 (it has no name), and B3/S45678 (Coral). We chose B3/S23 because it is the best-known automaton. B3678/S23 was chosen because it is similar to Life but it tends to produce patterns that are slightly denser than B3/S23. The rule B3/S45678 produces very dense patterns. Of course, given that there are 262,144 lifelike rules, there is a large degree of arbitrariness in our choice of these three particular rules.

# 3. Evolutionary Algorithms

Evolutionary algorithms were inspired by Darwin's theory of biological evolution (Spears, 1998; Simon, 2013). An evolutionary algorithm includes operations for mutation and selection, applied to a data store, just as biological evolution applies mutation and selection to DNA. This combination (mutation, selection, and a data store) describes a minimal instance of an evolutionary algorithm. Optional extras include growth, sexual recombination, segregation, genetic drift, coevolution, and symbiosis. We will use mutation, selection, growth, and a population of two-dimensional matrices for storing data.

## 3.1 Software: Golly and Python

In this article, we use the Golly Game of Life software, version 4.2 (Trevorrow and Rokicki, 2022), and the Python programming language, version 3.12. Golly was designed to be integrated with Python. Golly provides a flexible viewing environment for animating the growth of cellular automata on a computer screen. Python provides a programming environment that facilitates modifying and enhancing the capabilities of Golly. The Game of Life is included in the Golly software. Golly supports all 262,144 members of the Life family.

The evolutionary algorithm code is implemented in Python (Turney, 2024). The population of digital organisms is represented as a list of two-dimensional Python matrices. Only one organism appears on the Golly screen at a time. The organisms do not directly interact with each other. The organisms interact indirectly through their fitness scores. A two-dimensional Python matrix is sampled from the list of matrices and written on the Golly screen, where its fitness is then evaluated. The screen is then cleared for the next organism.

## 3.2 Mutation, Selection, and an Evolving Population of Patterns

Our playing field consists of a 60×60 grid of squares, displayed in Golly. The grid is a toroid (a doughnut). Imagine a 60×60 grid of squares, then wrap the grid into a tube by joining the top of the grid to the bottom of the grid. Next, join the left side of the tube to the right side of the tube. This creates a finite toroidal grid of 60×60 squares, with no borders. The grid looks like a square plane in the Golly screen, but it behaves like a toroid. For example, if a pattern moves across the right border of the square plane, it will reappear at the left border. If a pattern moves across the top of the grid, it will reappear at the bottom of the grid.

The advantage of a toroidal grid over a square grid is that there can be artifacts at the edges of a square grid, which tend to disrupt the patterns in the grid. An unbounded grid would also avoid artifacts, but it could increase computation time.

The playing field (the Golly toroid) contains only one organism at a time. The population of organisms is stored in a list of matrices (in Python), one matrix for each organism. Each matrix in the list of matrices is initially filled with white (zero) and then black (one) is randomly added to the matrices. The matrices are 60×60, but an organism is initially limited to the 30×30 center of the matrix. We call this initial 30×30 configuration a *seed*.

In our experiments, we begin with a population of 1,000 randomly generated seeds. To evaluate the fitness of a seed, it is copied from the Python list and then written into the Golly toroid. As the seed grows in the toroid over time, it will tend cover the 60×60 grid. We allow the seed to grow for 100 steps, at which point it is then an *adult*.

To evaluate the fitness of an adult, we compare the adult with a *target*. The target is an arbitrary 60×60 pattern that a human player creates. The fitness of an adult is calculated by measuring how well the adult matches with the target. The target could be any 60×60 pattern of black and white. The target is static.

We randomly sample two seeds from the population of 1,000 seeds. We score each of the two seeds. The seed with the higher score of the two is preserved in the population and the seed with the lower score is removed from the population. This reduces the population to 999. We then make a copy the seed with the higher score and mutate the copy. The new seed is 30×30, so it has 900 squares. The probability of mutation is set to 0.1, so the expected number of mutations in the new seed is 90 (900×0.1). The mutated copy is the offspring of the winning seed. It might be more fit than the parent or it might be less fit. The new seed brings the population back up to 1,000.

To score a seed, we first allow it to grow to adult size, by running the game for 100 steps. We then compare the adult with the target. The score is calculated by comparing each square in the adult with each corresponding square in the target. We start with a score of zero. We then examine each square in the adult (60×60 = 3,600 squares) and compare it with each corresponding square in the target (also 3,600 squares). If an adult square is black and the corresponding target square is black, the score increases by one point. If an adult square is black and the corresponding target square is white, the score decreases by one point. This fitness score rewards adults that put black squares on black targets and punishes adults that put black squares on white targets.

Note that the evolutionary algorithm cannot actually see the target pattern. It only receives a numerical fitness score that measures the similarity of the adult pattern to the target pattern. In effect, the evolutionary algorithm is blind to the positions of the squares. It only considers the total numbers of matching or mismatching colours.

As a seed grows into its adult form, Golly is responsible for managing the growth of the seed. Golly was designed to work closely with Python. When the seed becomes an adult, mutation and selection take place outside of Golly, running in an external Python program. The Python code is responsible for the evolution of new organisms, which are then delivered back to Golly.

There are several parameters that control the behaviour of the algorithm, listed in Table 1. These parameters are under the control of the external Python program. A seed is a *genome* and an adult is a *phenome*. Seeds and adults are both Python matrices. A seed grows into an adult in 100 steps, following the rules of the game, in Golly. Mutation and selection take place outside of Golly, in Python.

The parameters *prob_mutation* and *prob_selection* both range from 0 to 1. The target and the adult both have 60 × 60 = 3,600 cells. The parameters *prob_mutation* and *prob_selection* were tuned to maximize the growth of the seeds. The tuning with the best growth was *prob_mutation* = 0.1 and *prob_selection* = 0.6. The other parameters in Table 1 were not optimized.

| Parameters | Values | Description |
|---|---|---|
| *rule_name* | B3/S23 | B3/S23 (Life), B3678/S23 (no name), B3/S45678 (Coral). |
| *target_number* | 1 | A number for keeping track of the targets, from 1 to 5. |
| *population_size* | 1000 | The population size is constant: each death is followed by one birth. |
| *sample_size* | 40 | A random sample of part of the population (40 organisms). |
| *max_births* | 1,000,000 | The maximum number of births for running one experiment. |
| *num_steps* | 100 | The number of steps in the game, from seed to adult. |
| *prob_black* | 0.5 | The initial probability of a black cell in a seed. |
| *prob_white* | 0.5 | The initial probability of a white cell in a seed. |
| *prob_mutation* | 0.1 | The probability of switching white and black colours in a seed. |
| *prob_selection* | 0.6 | The probability of adding a new fit seed and dropping an unfit seed. |
| *seed_size* | 30 × 30 | A seed is made of 900 cells, arranged in a 30 × 30 square. |
| *adult_size* | 60 × 60 | An adult is made of 3,600 cells, arranged in a 60 × 60 square. |

Table 1: The parameters that determine how seeds and adults are formed.

To calculate the fitness of an organism (a matrix), we scan through the 3,600 cells and compare each target cell with each adult cell in the same relative position (see Figure 1 below). We start with a score of zero. Each time both the adult cell and the corresponding target cell are black, we add one point to the score (because the adult cell hit the target). Each time the adult cell is black and the corresponding target cell is white, we subtract one point from the score (because the adult cell missed the target). The total, after all 3,600 cells are scanned, is the fitness score of the current organism.

## 4. Playing the Games

Figure 1 summarizes the results of running the three rules: B3/S23, B3678/S23, and B3/S45678. All three share the same target. Each seed is the result of using an evolutionary algorithm to optimize fitness. Fitness is determined by comparing the target with the adult form of the corresponding seed.

The evolutionary algorithm runs for one million births. This might seem like a large number, but it only takes about two hours to run one million births on a desktop computer, and many biological species undergo far more than one million births. The population size is fixed at 1000.

| | | |
|---|---|---|
| 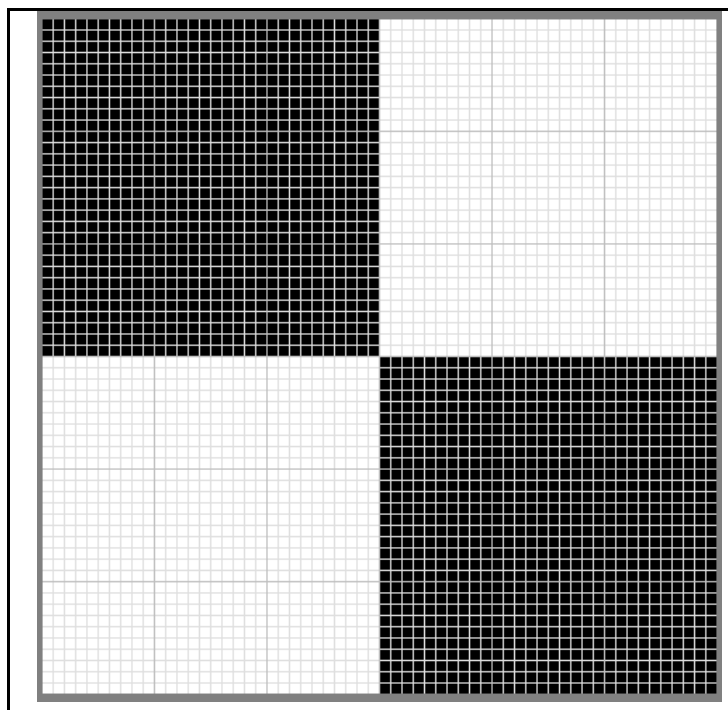<br>Target 1, 60 × 60 | 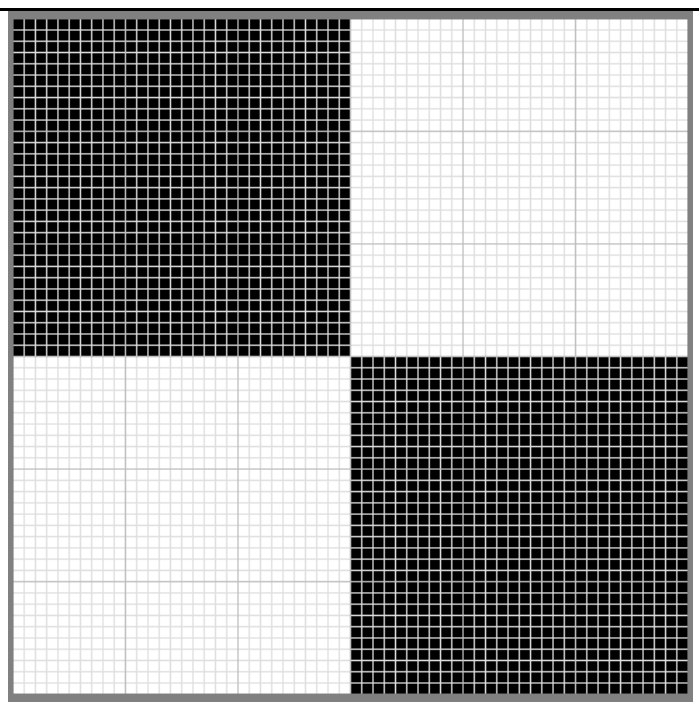<br>Target 1, 60 × 60 | 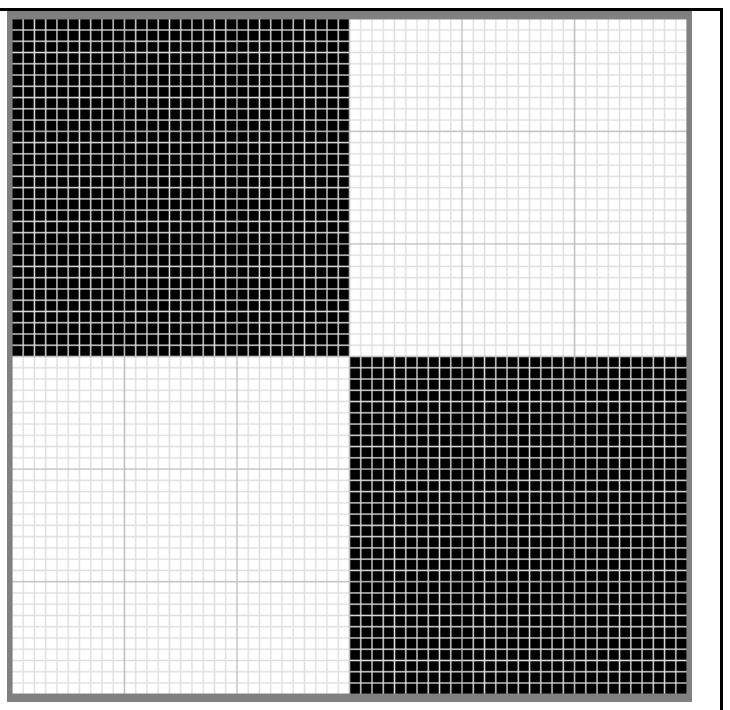<br>Target 1, 60 × 60 |
| 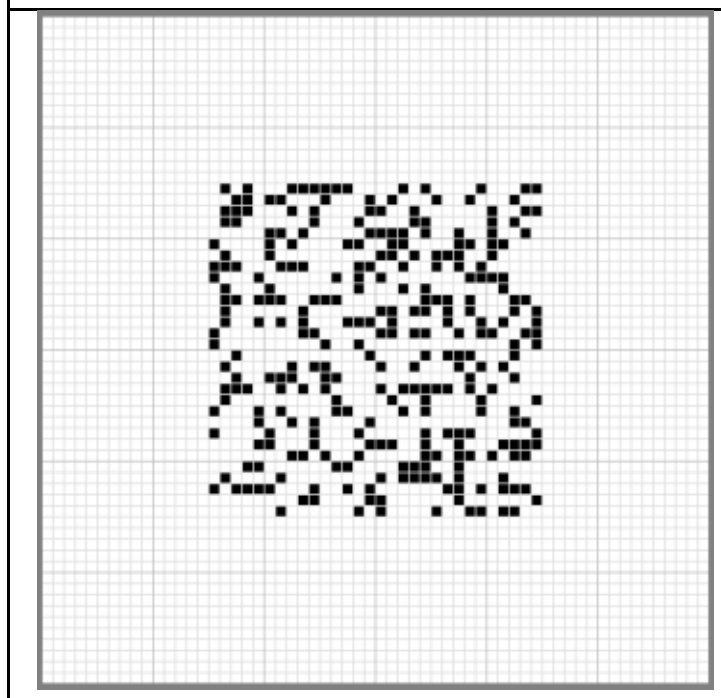<br>Seed 1, B3/S23<br>289 black squares, 30 × 30 | 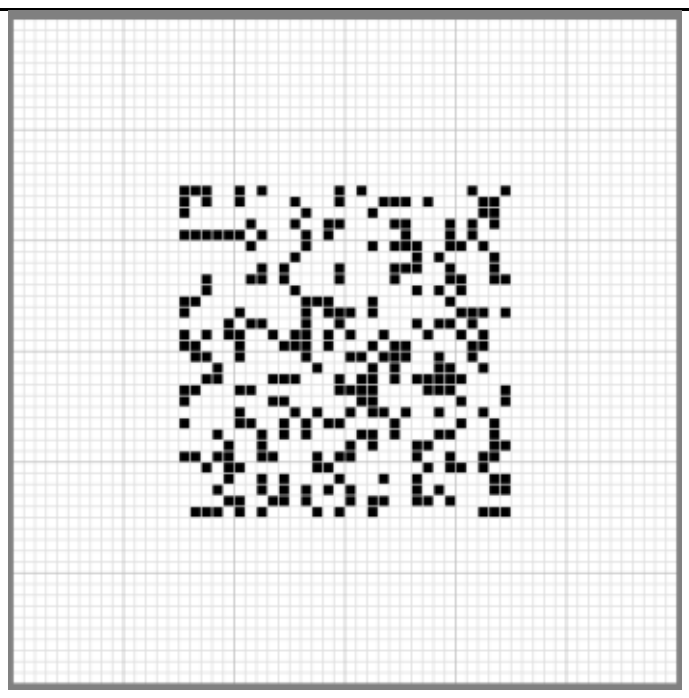<br>Seed 1, B3678/S23<br>284 black squares, 30 × 30 | 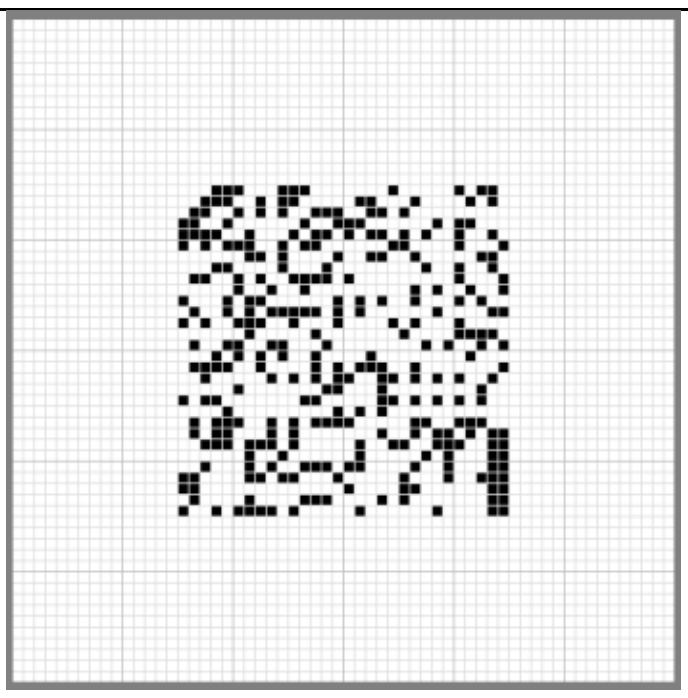<br>Seed 1, B3/S45678<br>304 black squares, 30 × 30 |
| 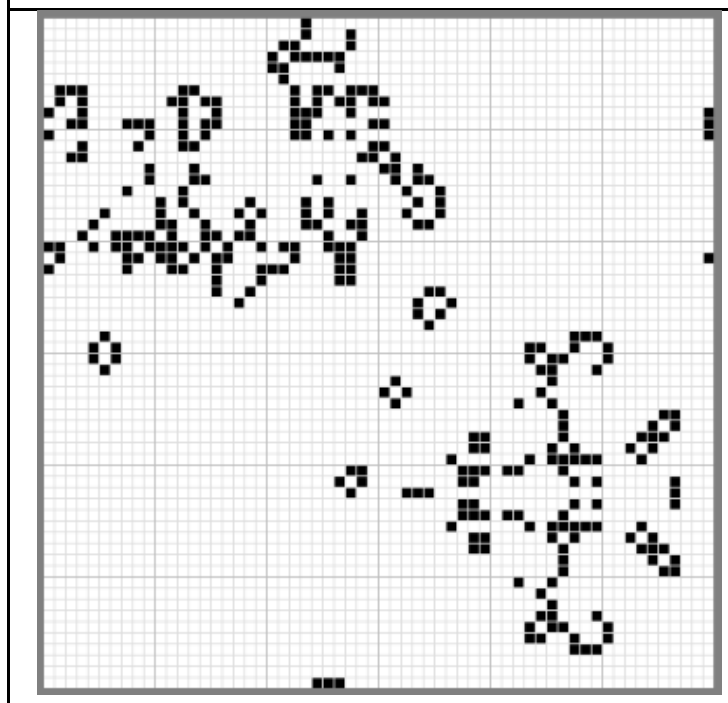<br>Adult 1, B3/S23<br>323 black squares, 60 × 60 | 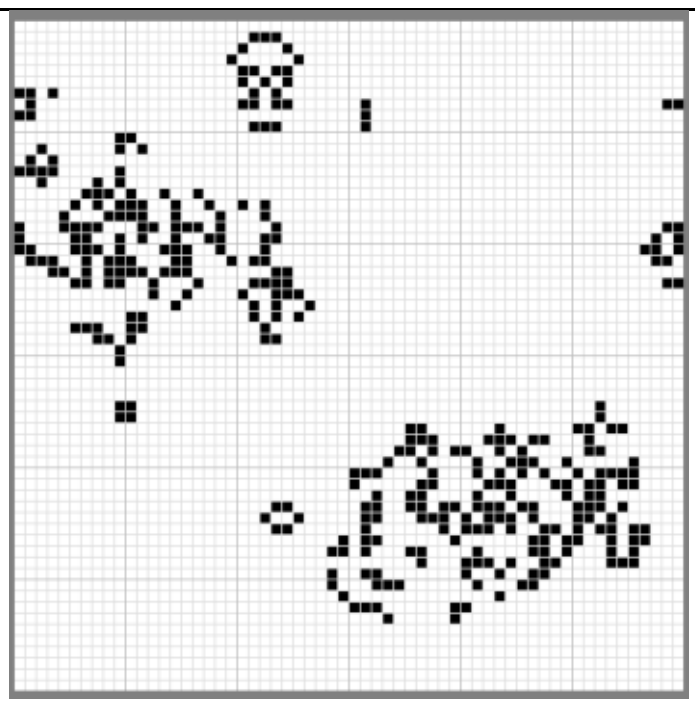<br>Adult 1, B3678/S23<br>352 black squares, 60 × 60 | 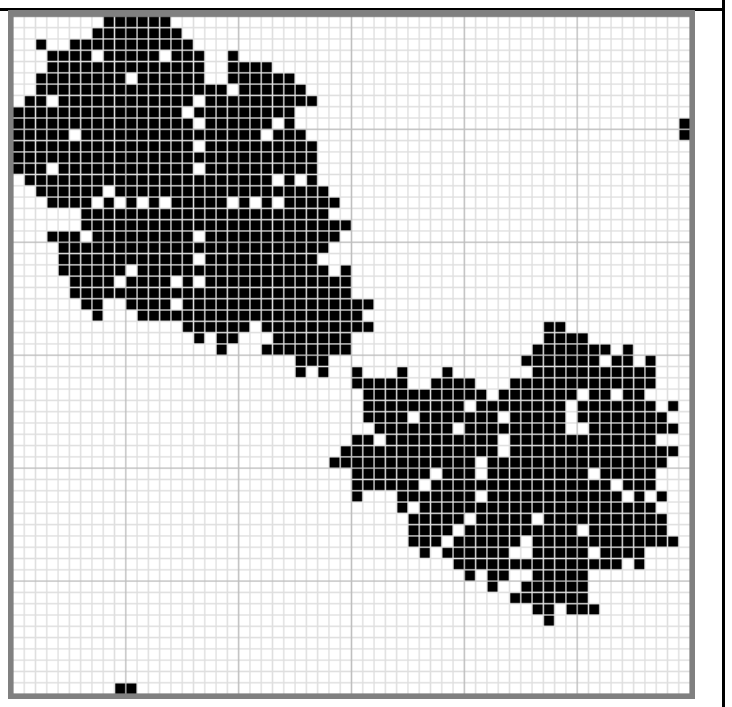<br>Adult 1, B3/S45678<br>1,071 black squares, 60 × 60 |

Figure 1: The Game of Life (B3/S23) struggles to match Target 1. B3678/S23 has slightly more success. B3/S45678 (Coral) gives a reasonable approximation of Target 1.

From one point of view, *engineering* cellular automata is quite different from *evolving* cellular automata. However, a closer look suggests the two approaches are not all that different. Engineering new cellular automata often involves significant trial and error experimentation. Evolving new cellular automata requires automating that experimentation.

| | | |
|---|---|---|
| 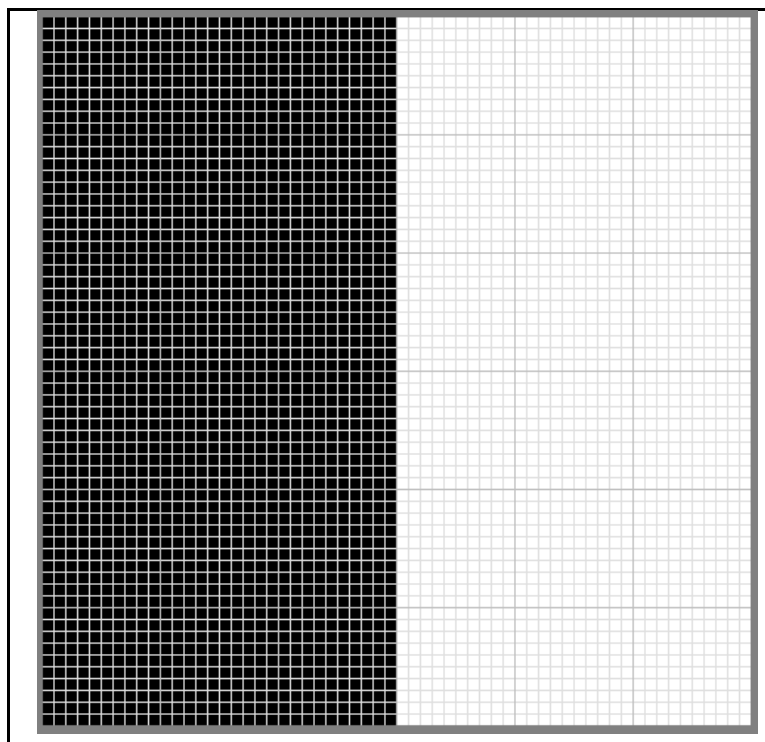<br>Target 2, 60 × 60 | 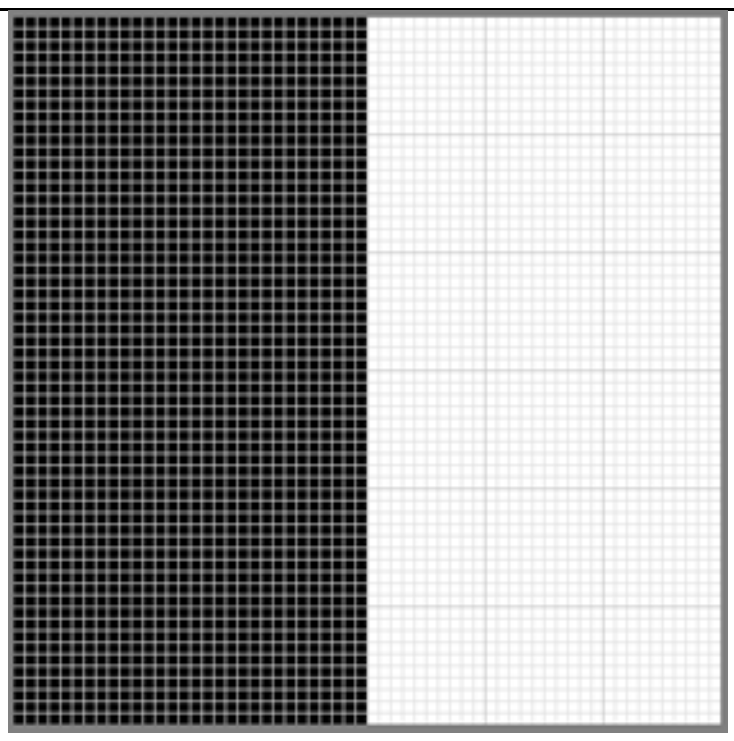<br>Target 2, 60 × 60 | 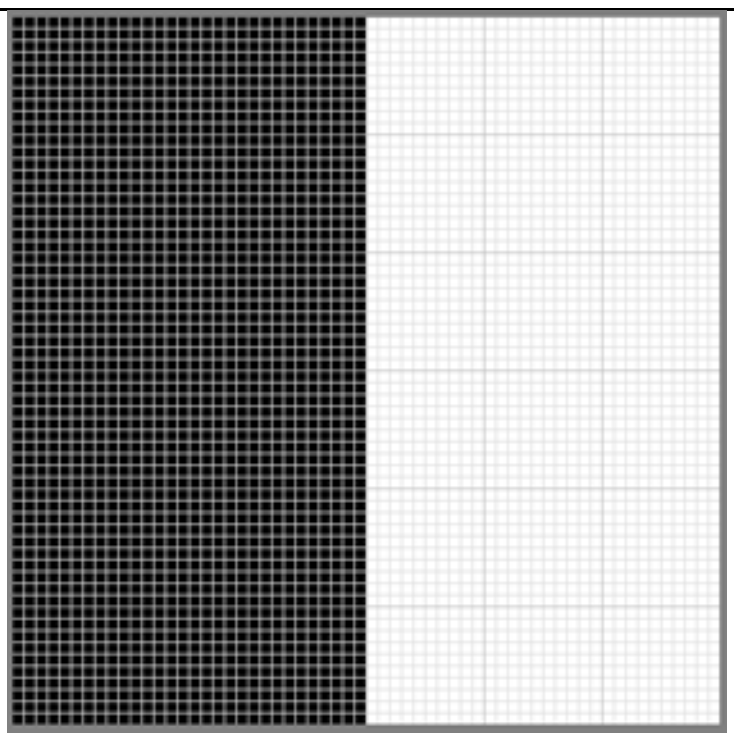<br>Target 2, 60 × 60 |
| 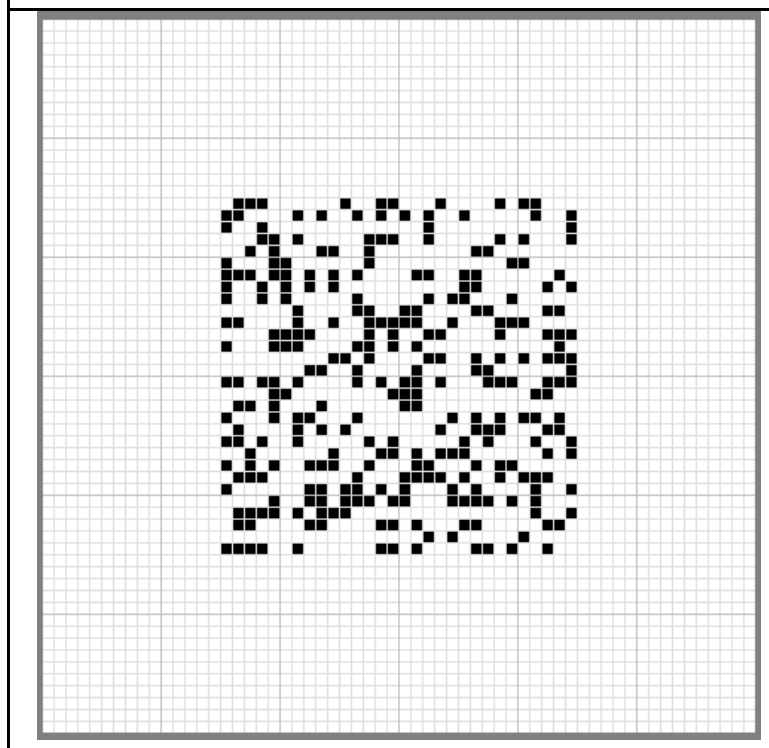<br>Seed 2, B3/S23<br>295 black squares, 30 × 30 | 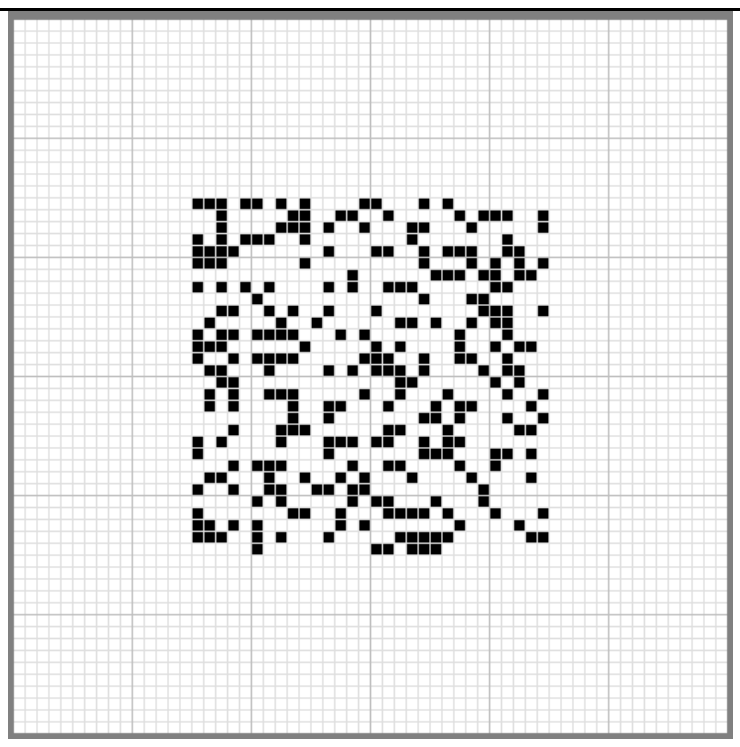<br>Seed 2, B3678/S23<br>283 black squares, 30 × 30 | 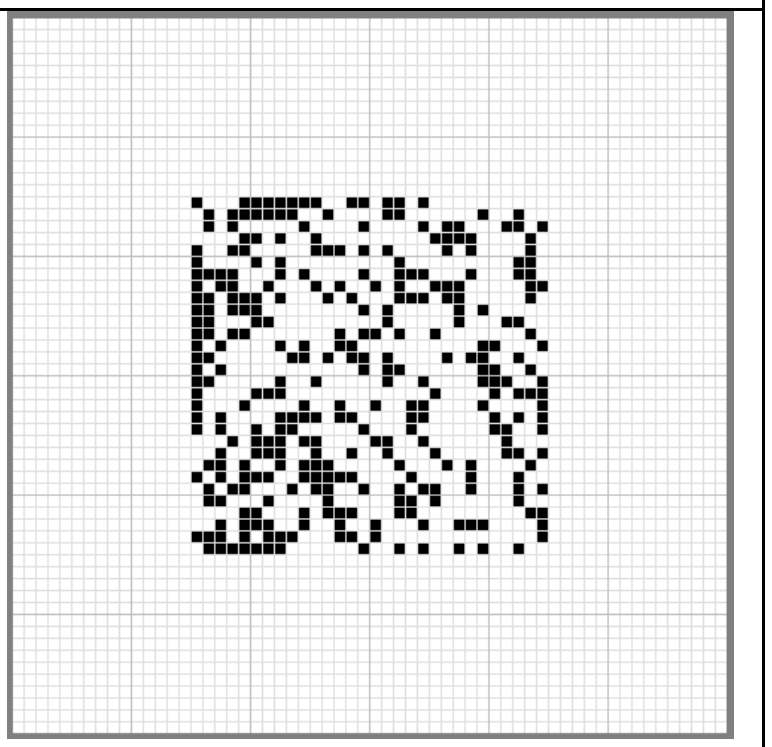<br>Seed 2, B3/S45678<br>320 black squares, 30 × 30 |
| 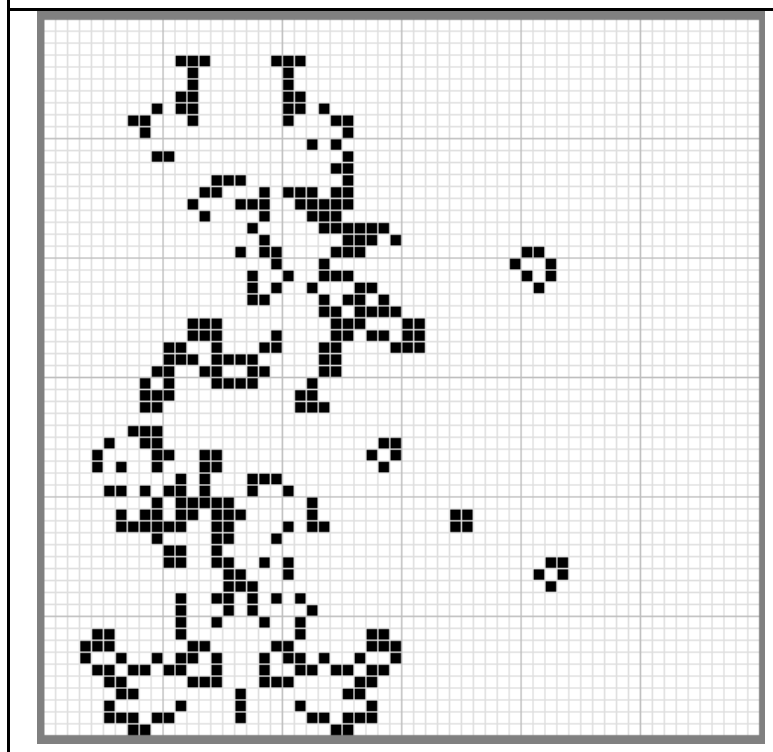<br>Adult 2, B3/S23<br>352 black squares, 60 × 60 | 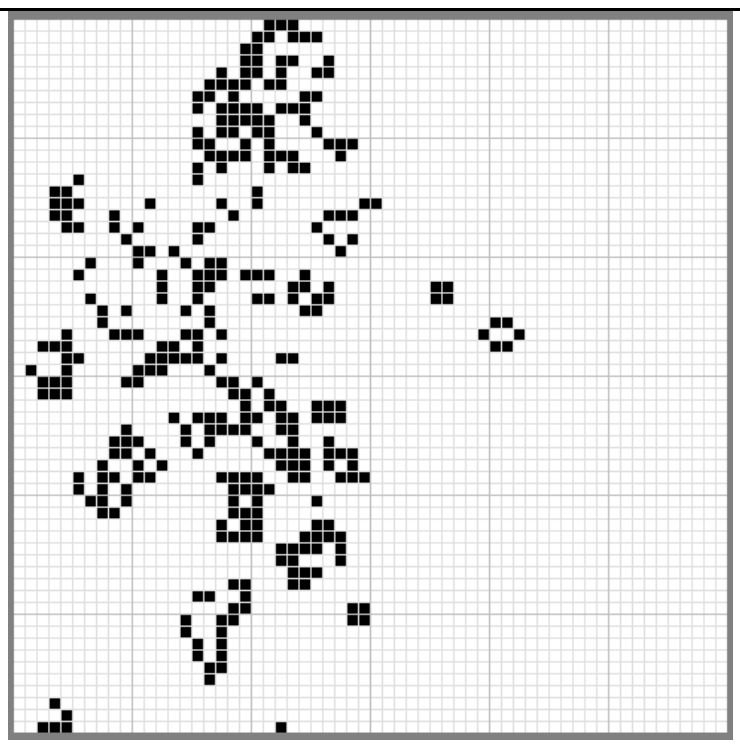<br>Adult2, B3678/S23<br>335 black squares, 60 × 60 | 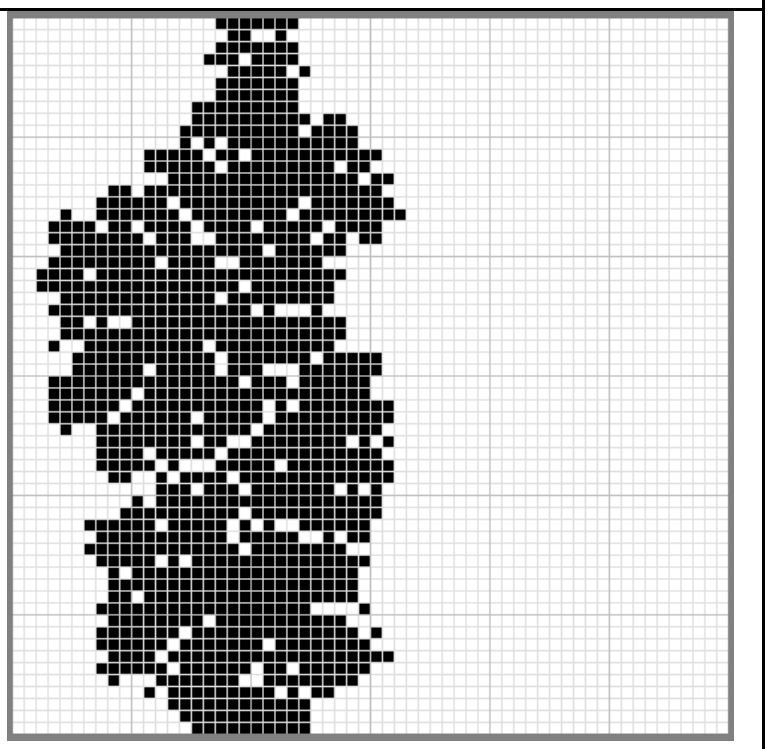<br>Adult2, B3/S45678<br>1,097 black squares, 60 × 60 |

Figure 2: B3/S45678 (Coral) works well with a single, thick black bar as the target.

The black bar in Target 2 is thicker than the black bar in Target 3, but general appearance of the two corresponding Adults is much the same. The variation between them is minor.

| | | |
|---|---|---|
| 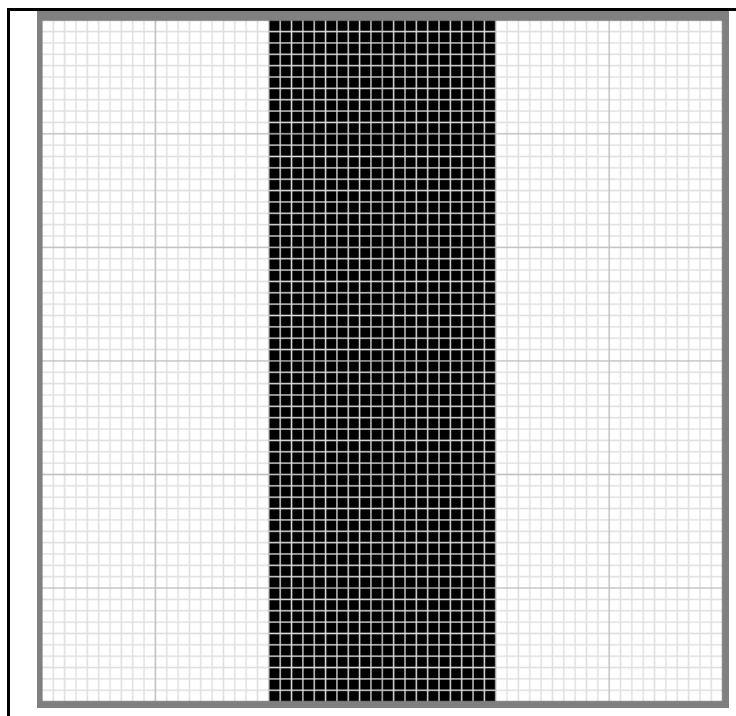<br>Target 3, 60 × 60 | 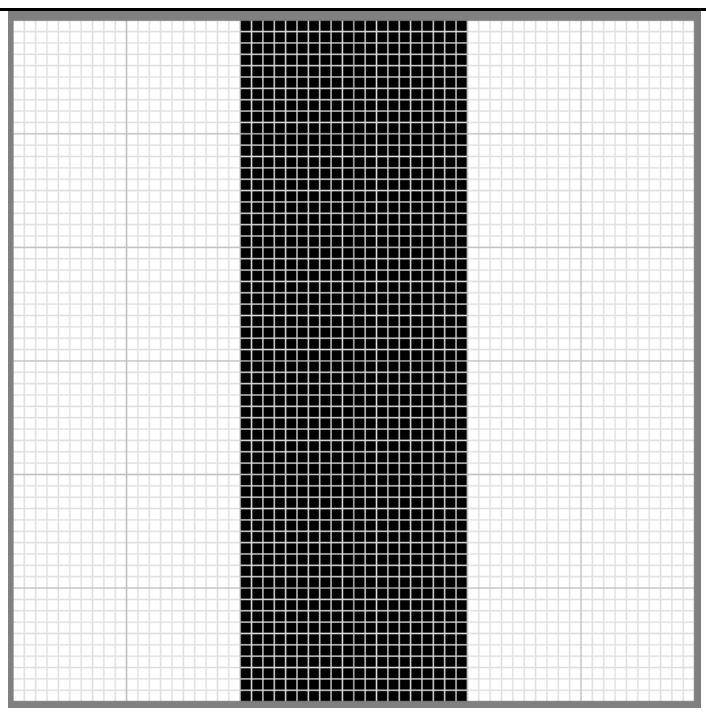<br>Target 3, 60 × 60 | 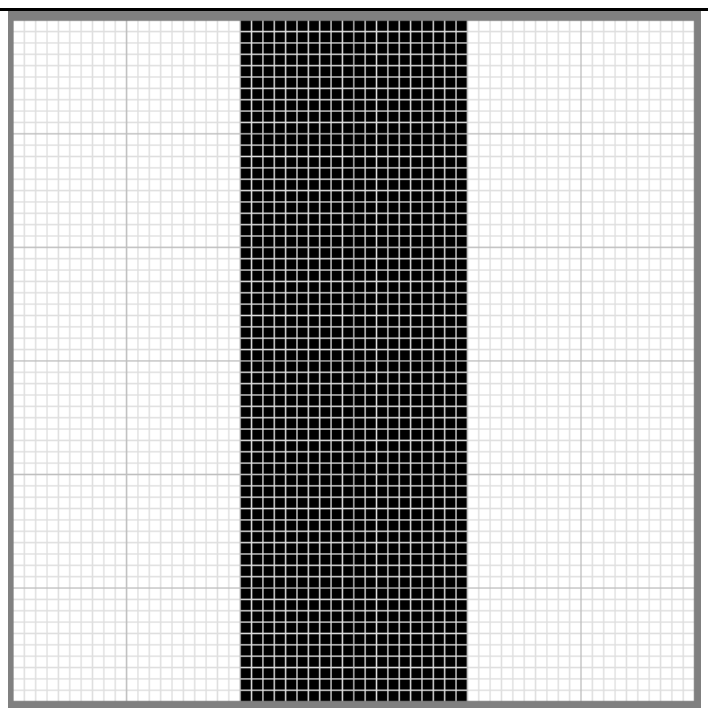<br>Target 3, 60 × 60 |
| 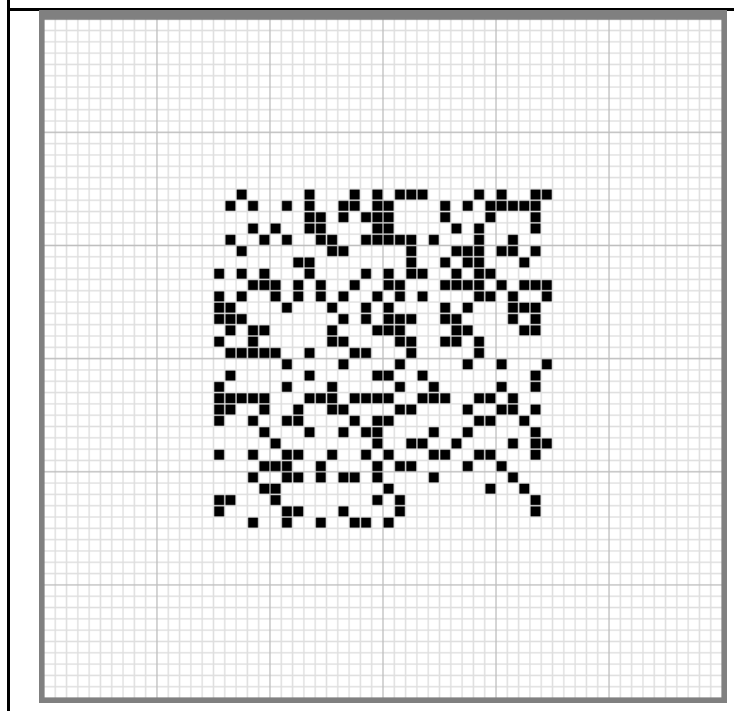<br>Seed 3, B3/S23<br>295 black squares, 30 × 30 | 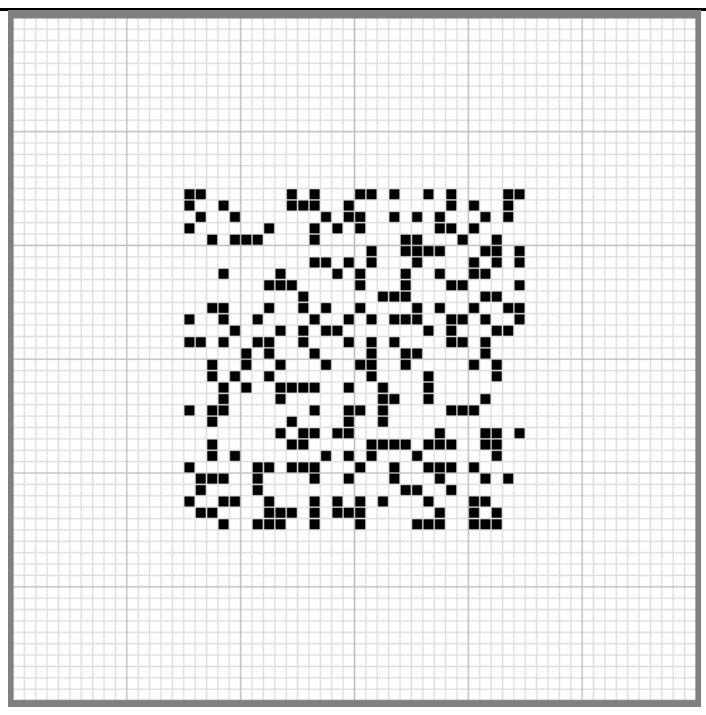<br>Seed 3, B3678/S23<br>271 black squares, 30 × 30 | 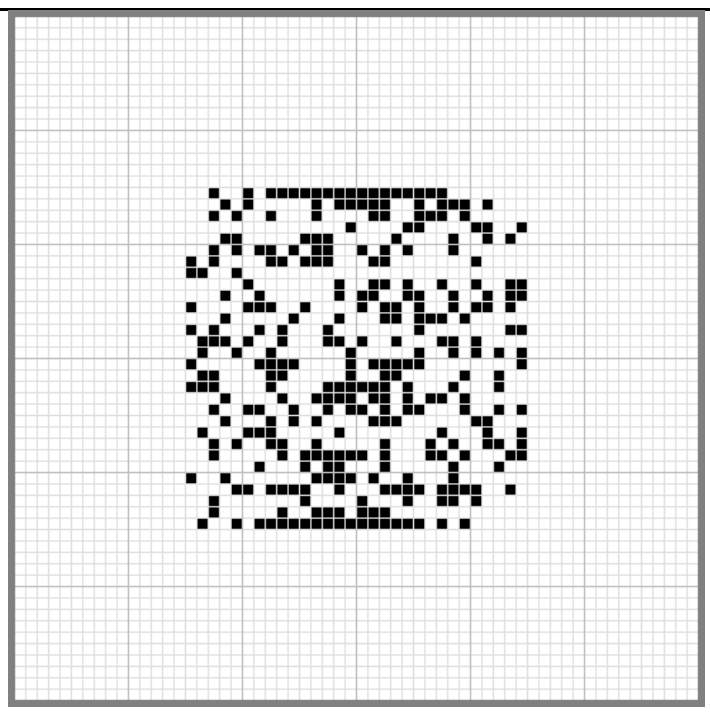<br>Seed 3, B3/S45678<br>318 black squares, 30 × 30 |
| 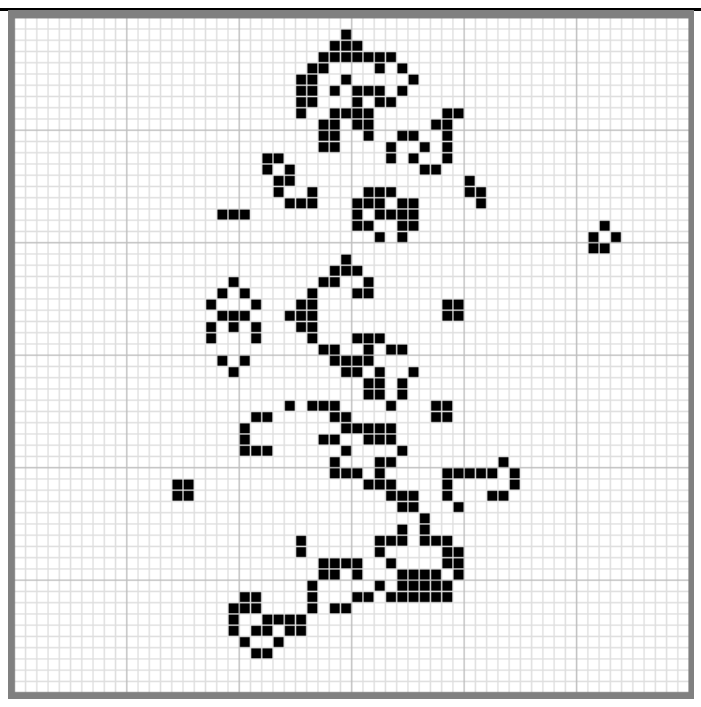<br>Adult 3, B3/S23<br>291 black squares, 60 × 60 | 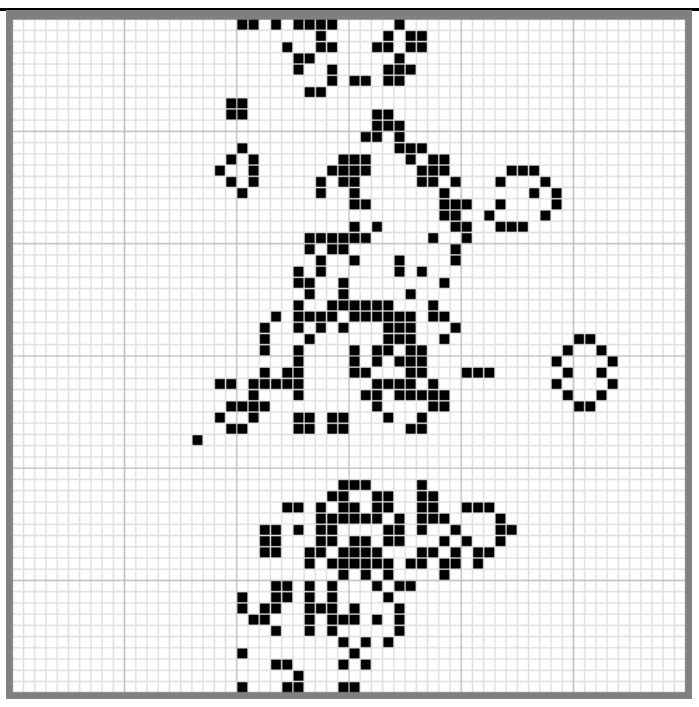<br>Adult 3, B3678/S23<br>380 black squares, 60 × 60 | 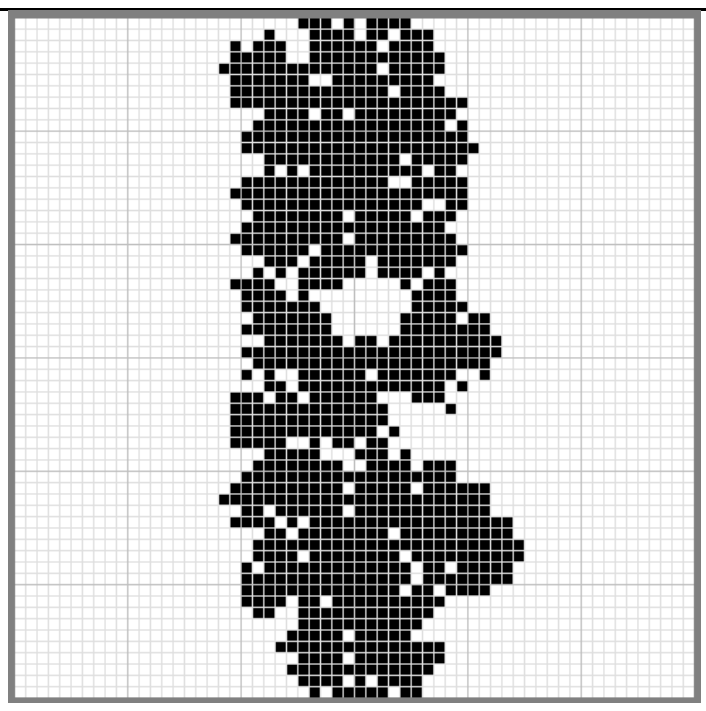<br>Adult 3, B3/S45678<br>979 black squares, 60 × 60 |

Figure 3: B3/S45678 (Coral) is consistently the densest adult.

| | | |
|---|---|---|
| 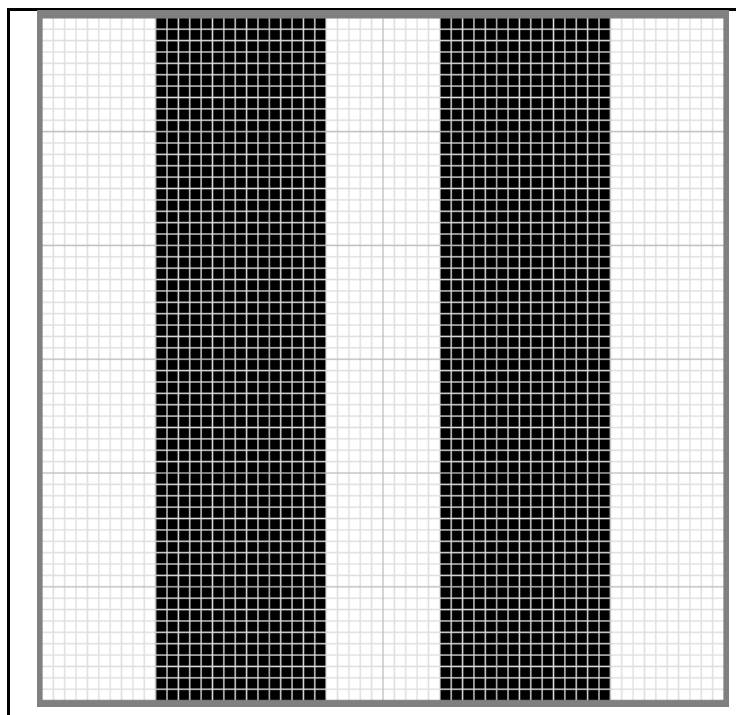<br>Target 4, 60 × 60 | 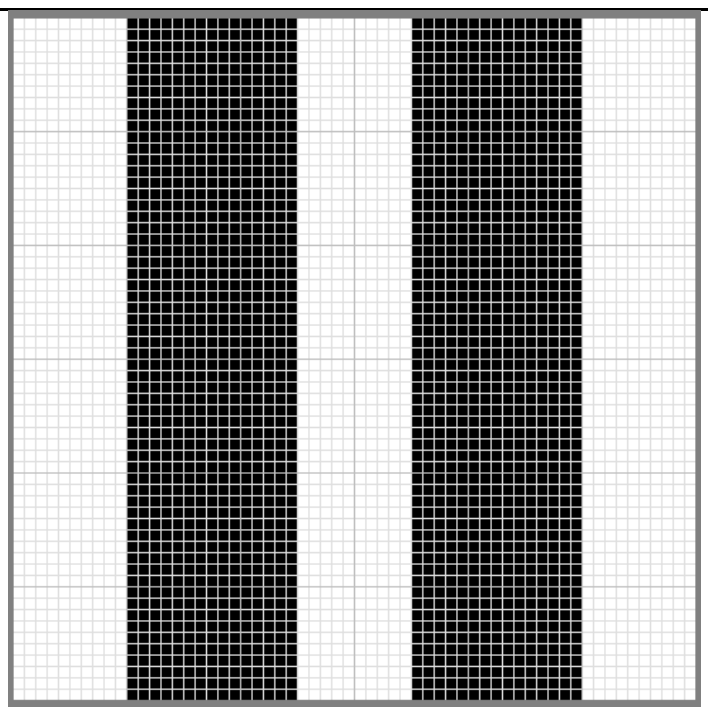<br>Target 4, 60 × 60 | 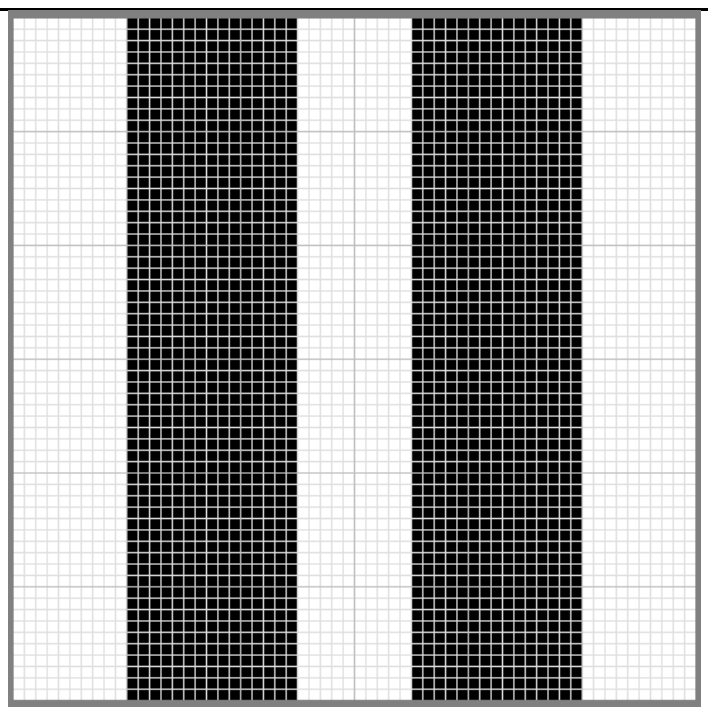<br>Target 4, 60 × 60 |
| 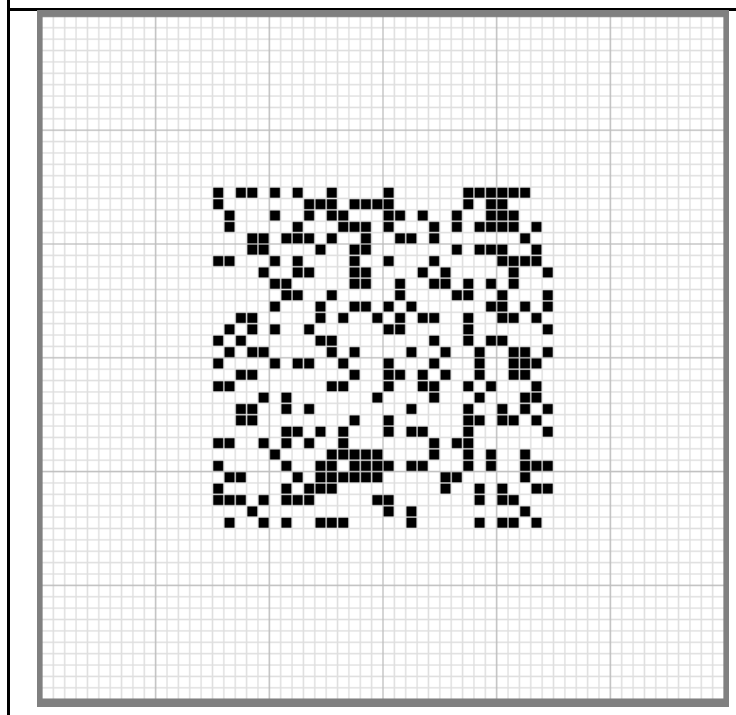<br>Seed 4, B3/S23<br>305 black squares, 30 × 30 | 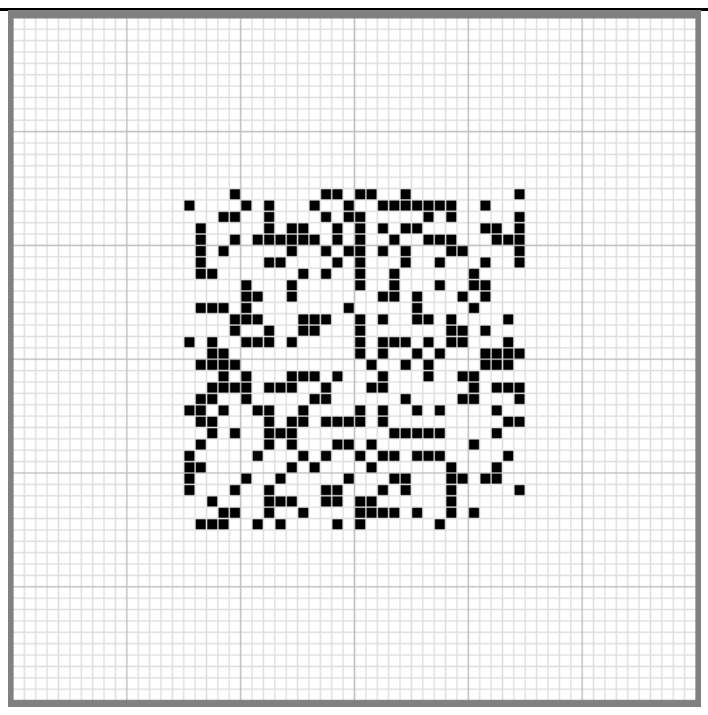<br>Seed 4, B3678/S23<br>300 black squares, 30 × 30 | 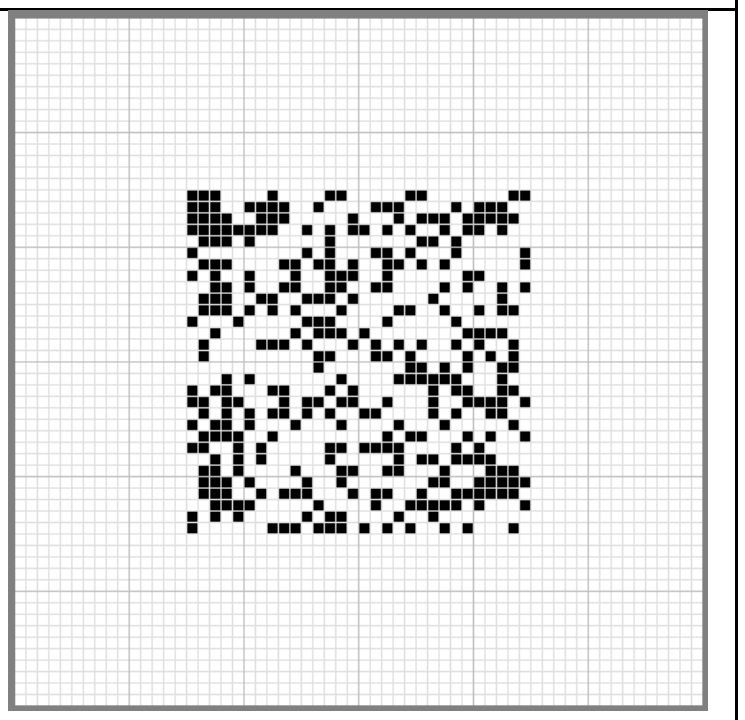<br>Seed 4, B3/S45678<br>351 black squares, 30 × 30 |
| 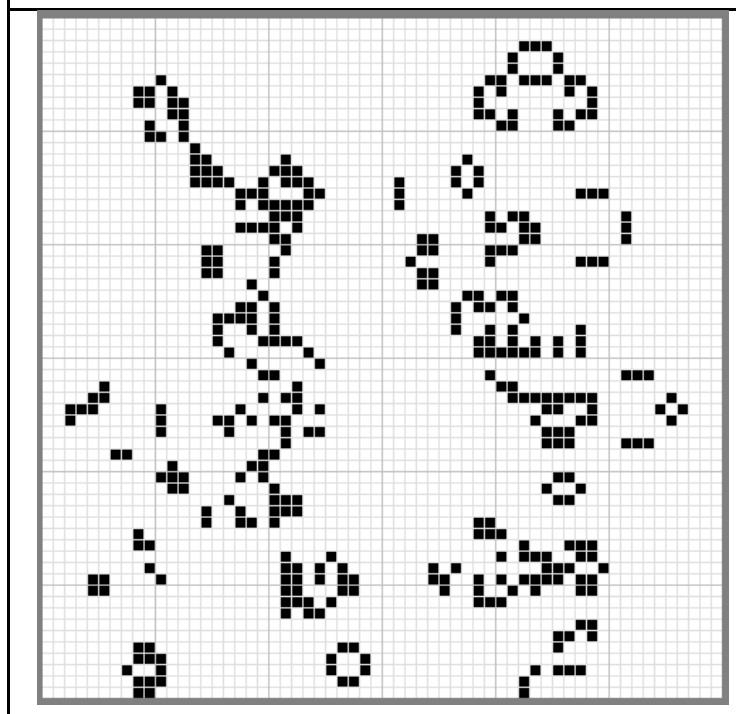<br>Adult 4, B3/S23<br>382 black squares, 60 × 60 | 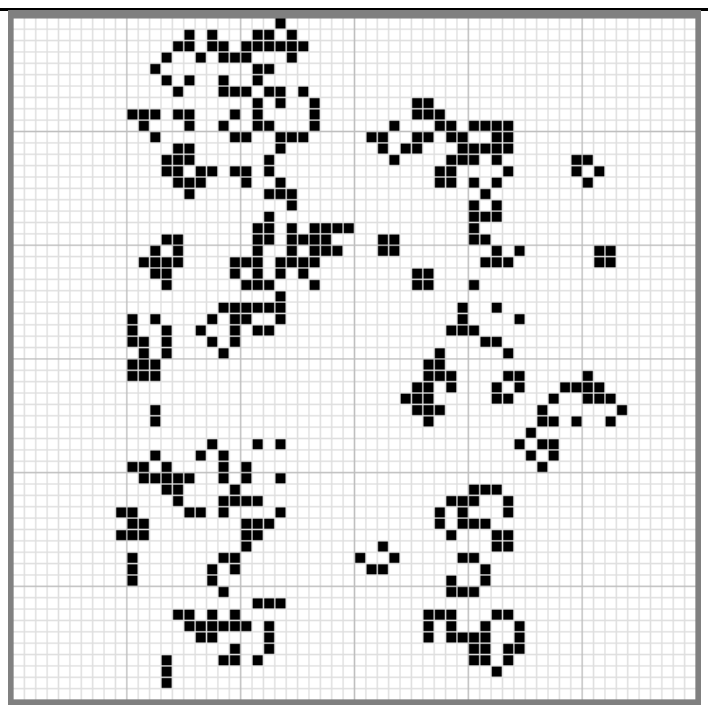<br>Adult 4, B3678/S23<br>422 black squares, 60 × 60 | 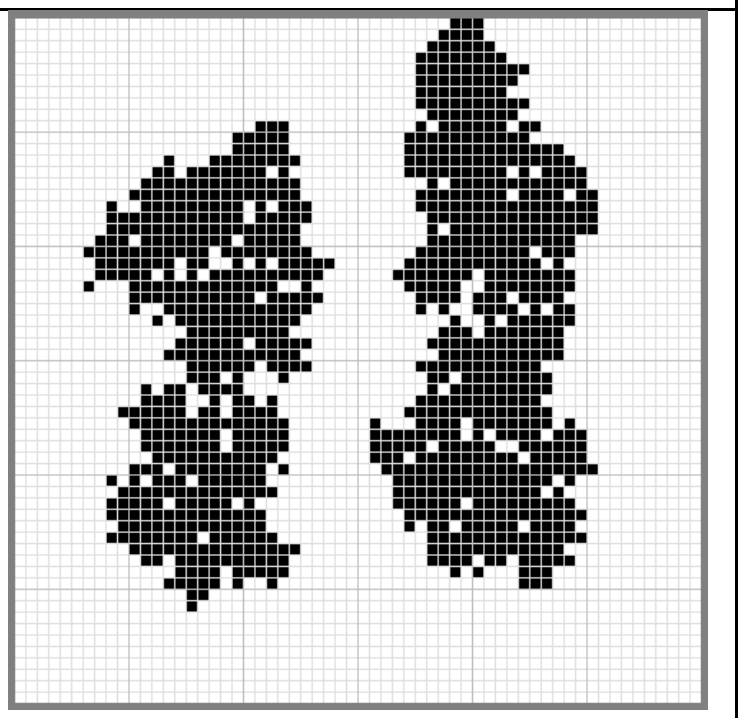<br>Adult 4, B3/S45678<br>1,074 black squares, 60 × 60 |

Figure 4: Adult 4, B3/S45678, seems to be struggling with the white space above and below the black patterns. The narrow gap between the two black blobs may have disrupted their growth. Note that Adult 4, B3/S23, and Adult 4, B3678/S23, were able to stretch out further than Adult 4, B3/S45678.

| | | |
|---|---|---|
| 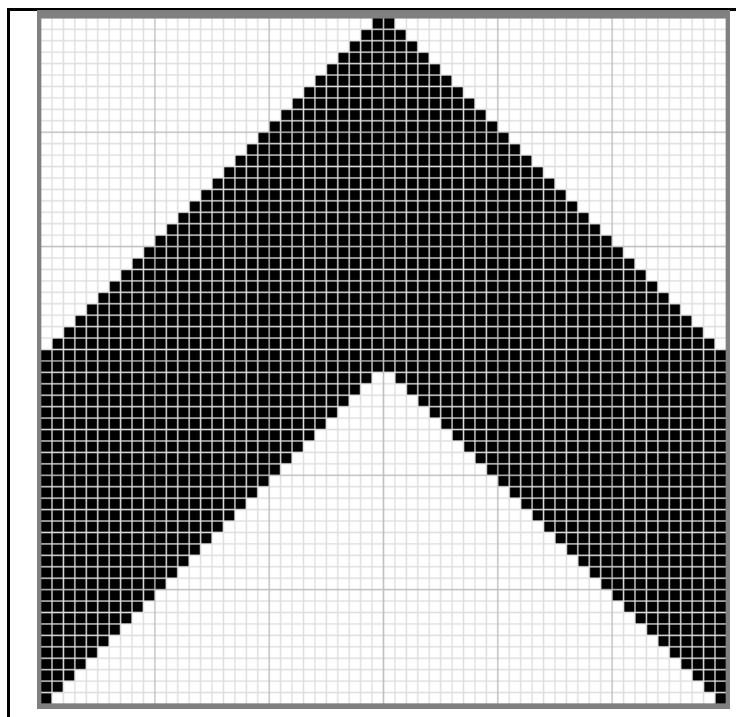 Target 5, 60 × 60 | 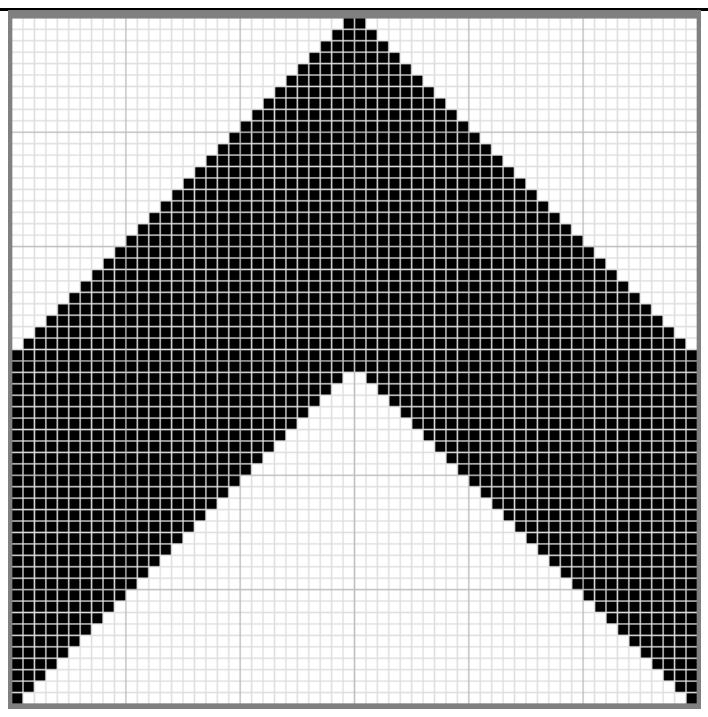 Target 5, 60 × 60 | 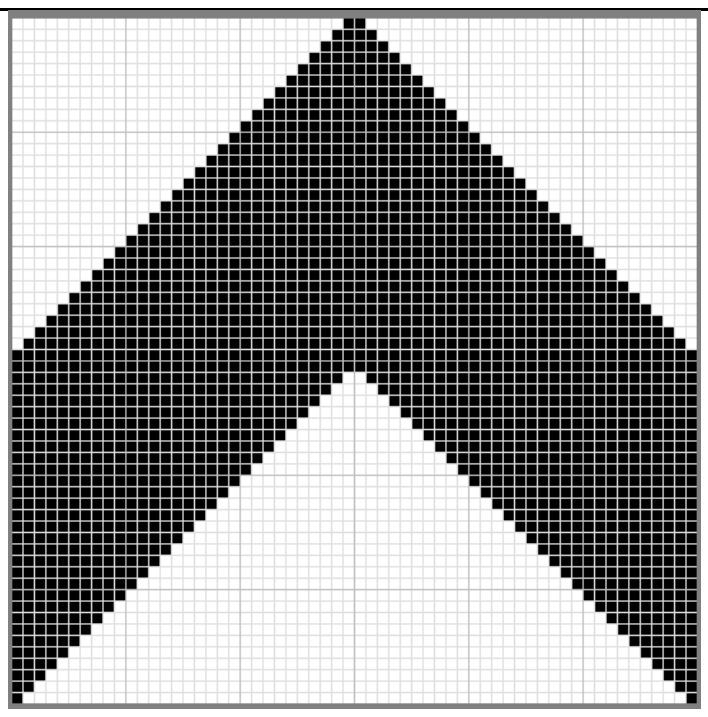 Target 5, 60 × 60 |
| 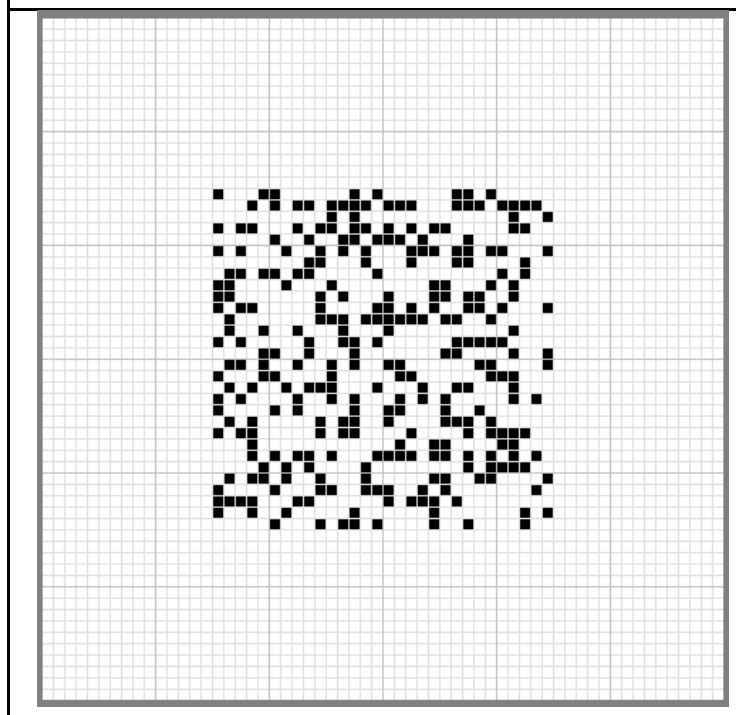 Seed 5, B3/S23<br>299 black squares, 30 × 30 | 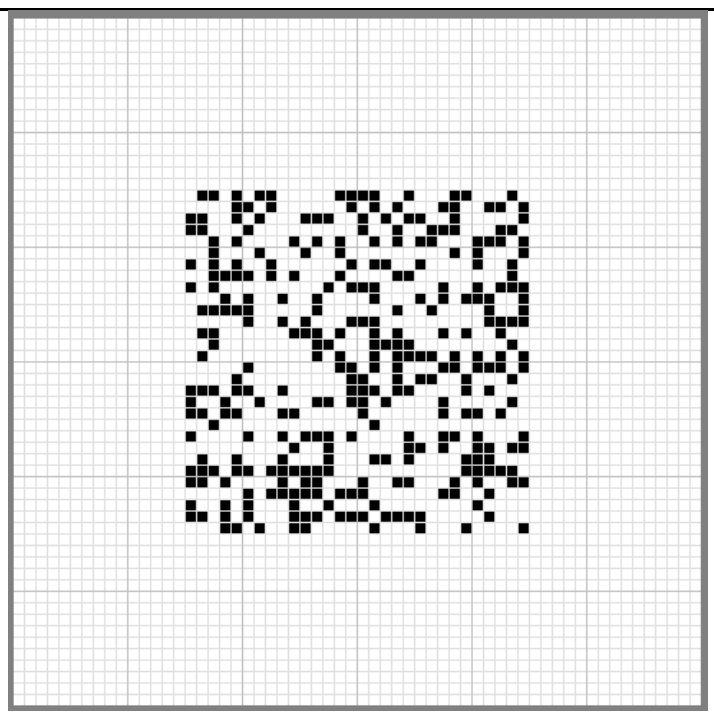 Seed 5, B3678/S23<br>318 black squares, 30 × 30 | 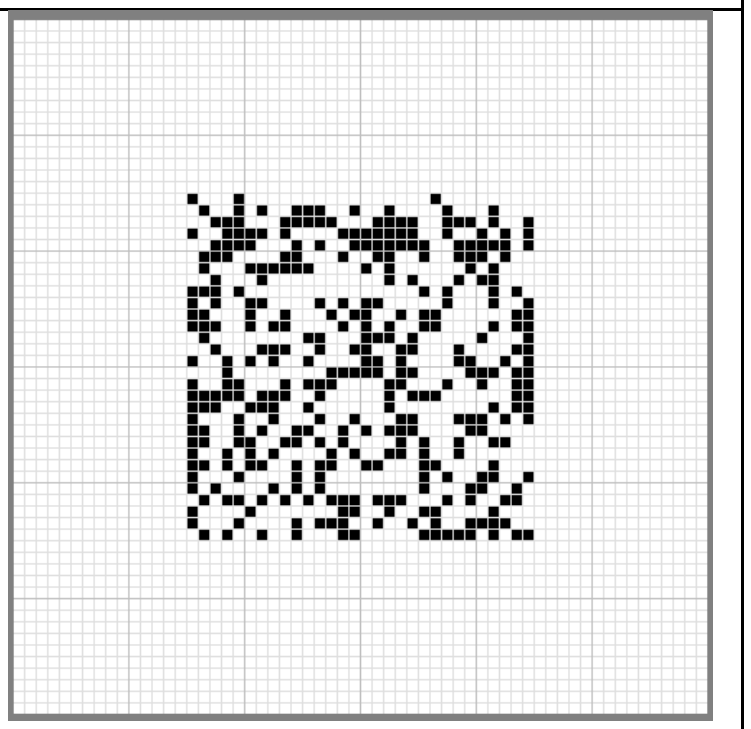 Seed 5, B3/S45678<br>357 black squares, 30 × 30 |
| 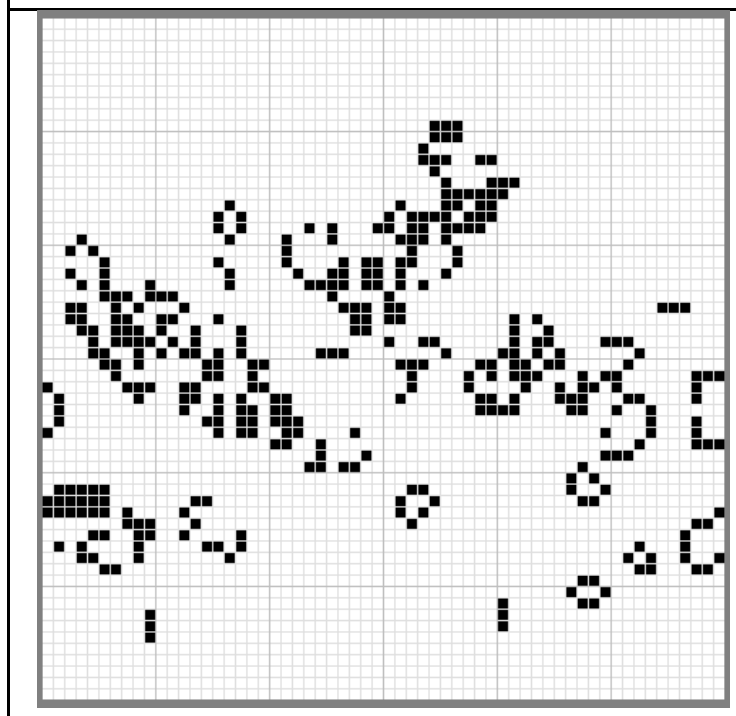 Adult 5, B3/S23<br>376 black squares, 60 × 60 | 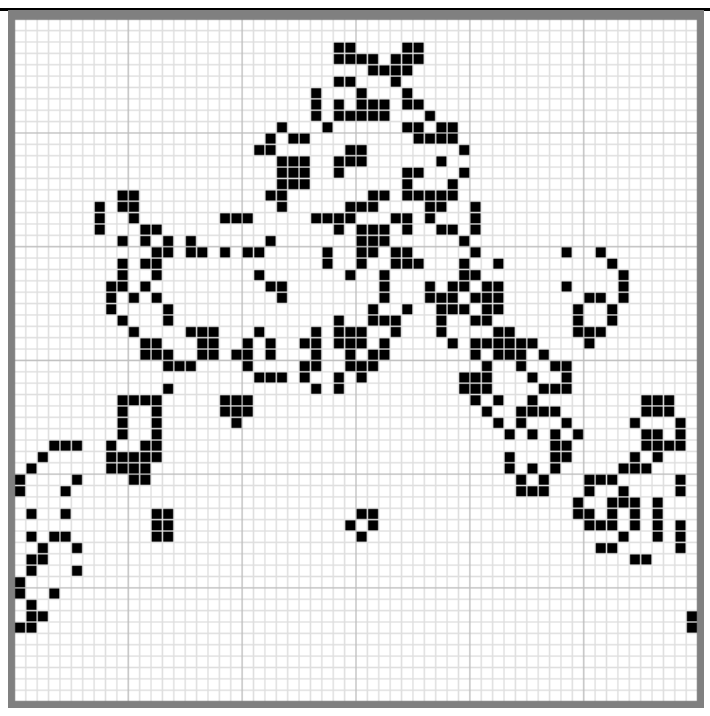 Adult 5, B3678/S23<br>446 black squares, 60 × 60 | 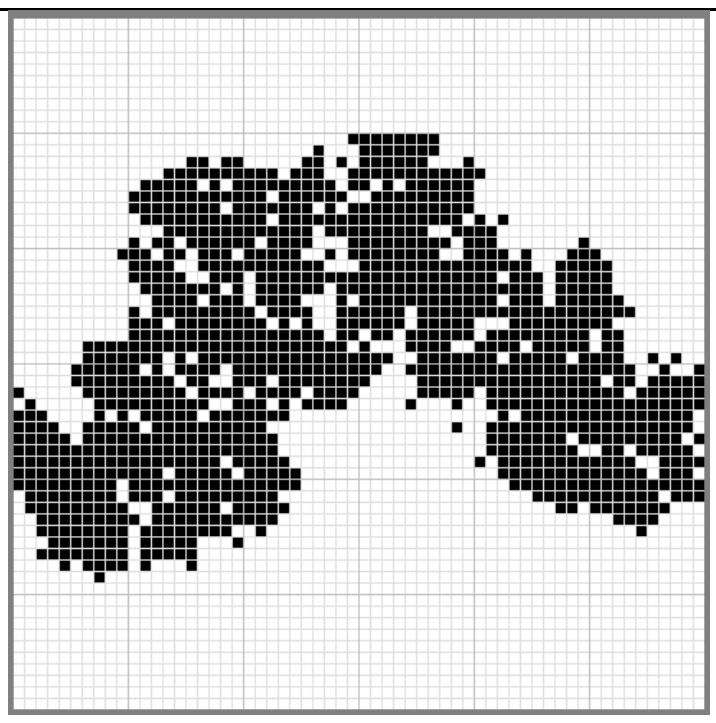 Adult 5, B3/S45678<br>1,161 black squares, 60 × 60 |

Figure 5: The diagonal lines in Target 5 appear to be more challenging than the horizontal and vertical lines in the first four figures. Adult 5, B3678/S23, captures the upside-down shape well, perhaps better than Adult 5, B3/S45678.

It takes 100 steps for a seed to become an adult. It is not practical to show all the steps here, but we can show the growth of the seed in a series of small jumps. Below we show B3/S45678, growing from seed to adult in twelve jumps.

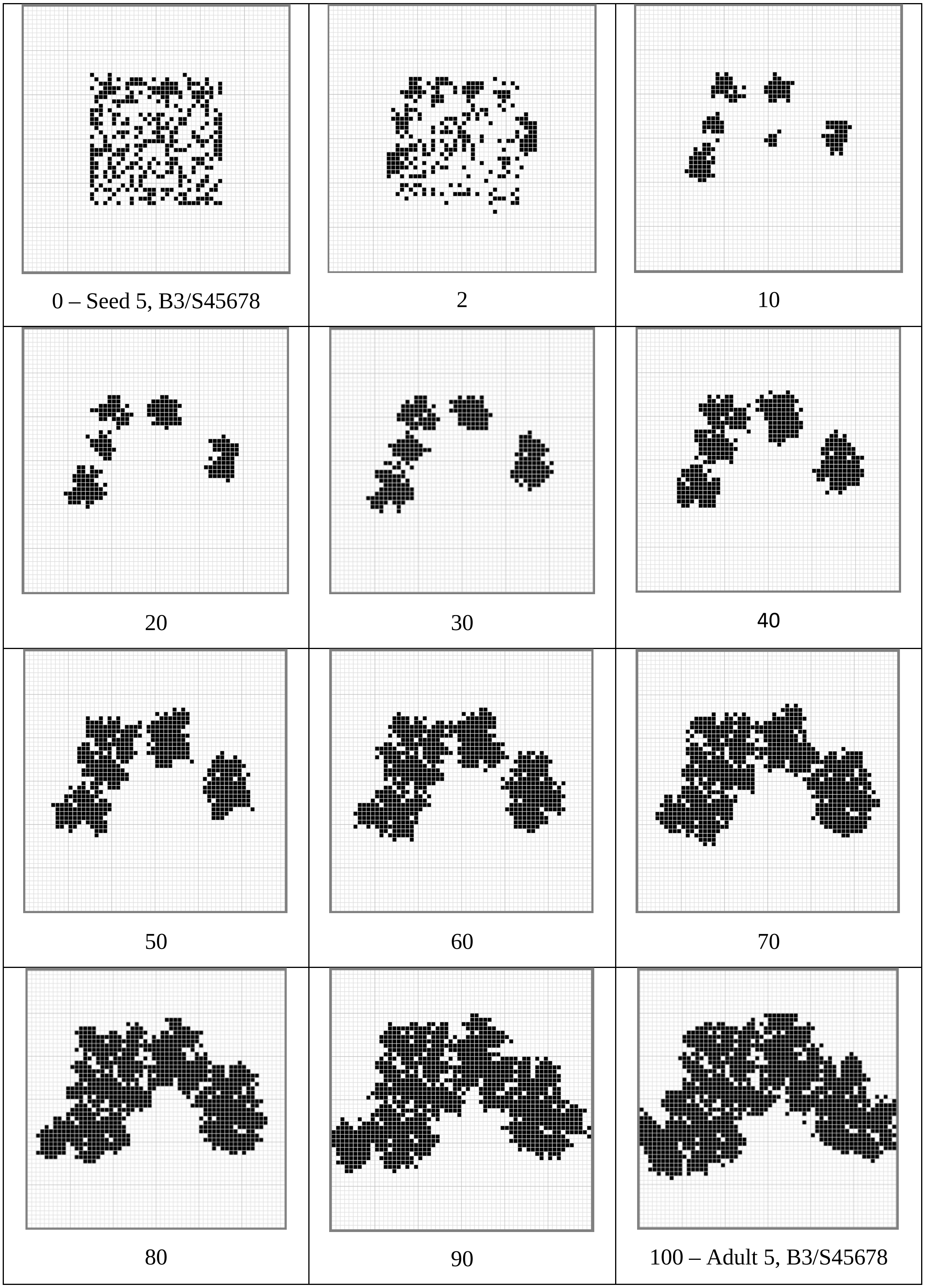

Figure 6: B3/S45678 grows from a seed (step 0) to an adult (step 100).

In Section 3.2, in Table 1, we have *prob_mutation* = 0.1, *prob_selection* = 0.6, and *max_births* = 1,000,000. Here in Figure 7, we examine the impact of setting *prob_mutation* or *prob_selection* to zero.

| | | |
|---|---|---|
| 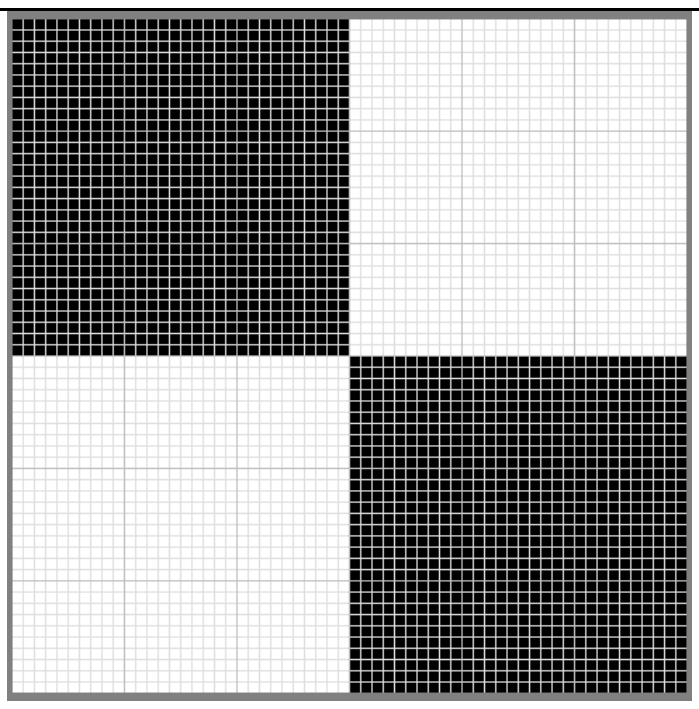<br>Target 1, 60 × 60 | 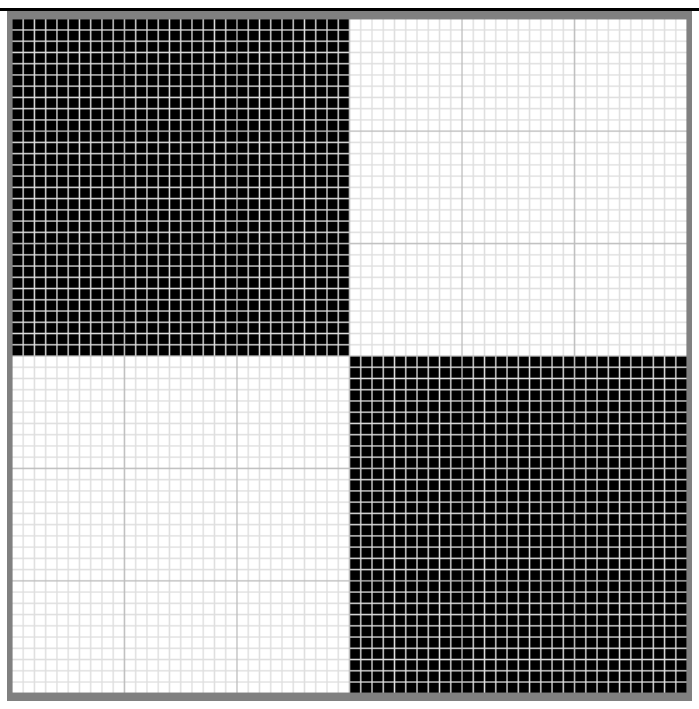<br>Target 1, 60 × 60 | 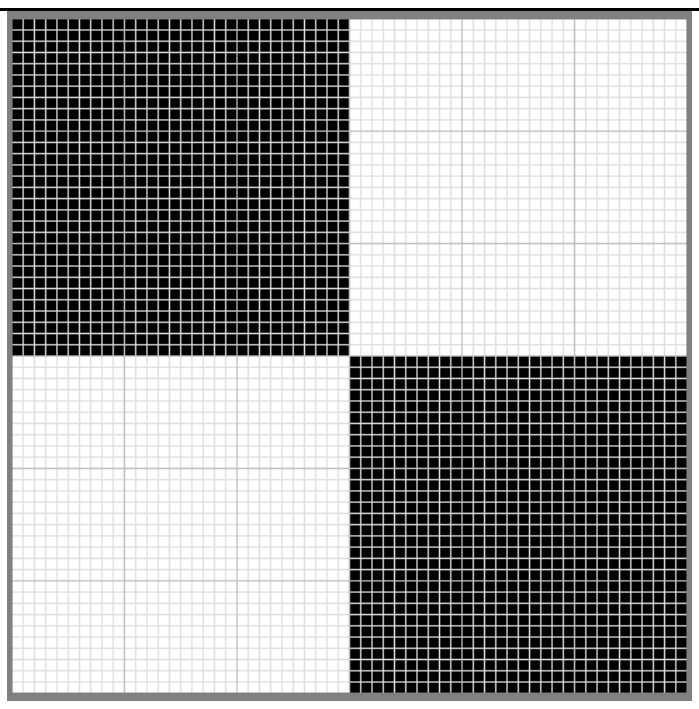<br>Target 1, 60 × 60 |
| 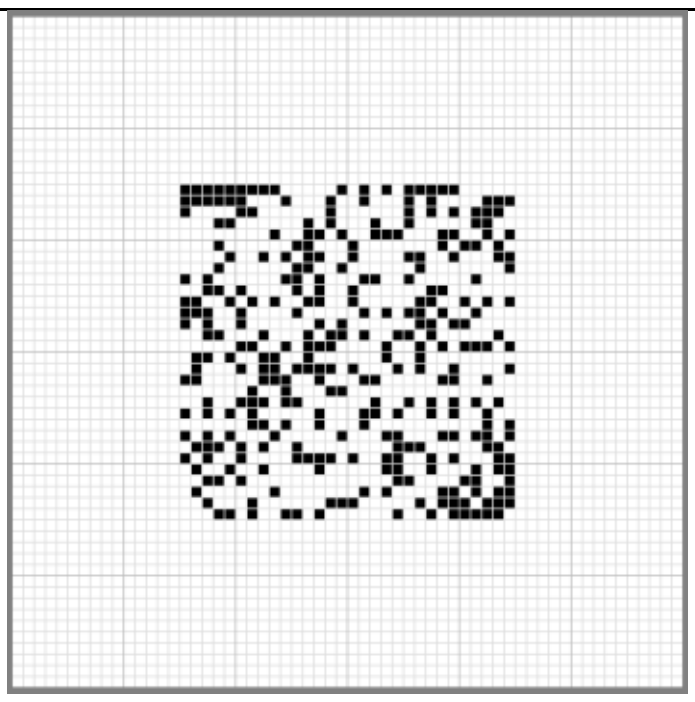<br>Seed 1, B3/S45678 | 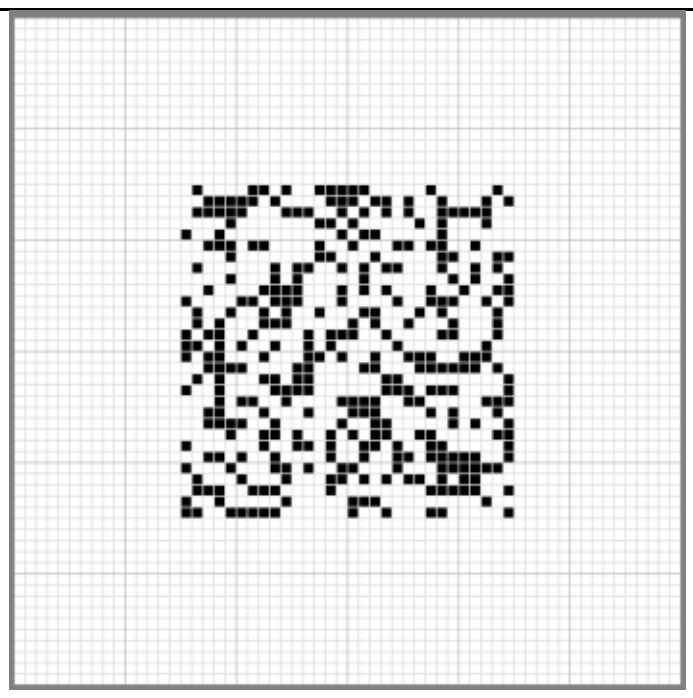<br>Seed 1, B3/S45678 | 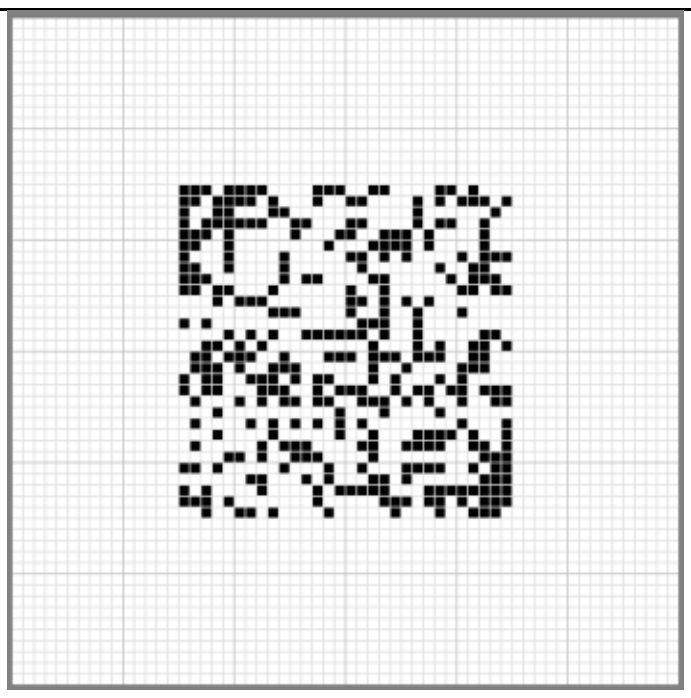<br>Seed 1, B3/S45678 |
| 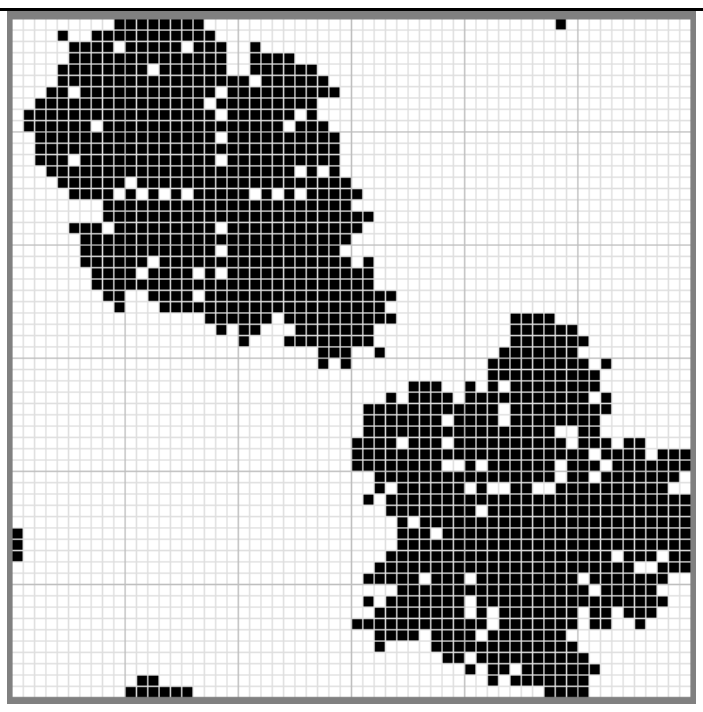<br>Adult 1, B3/S45678<br>*prob_mutation* = 0.1<br>*prob_selection* = 0.6<br>fitness = 1,106<br>black on target = 1,169<br>black off target = 63 | 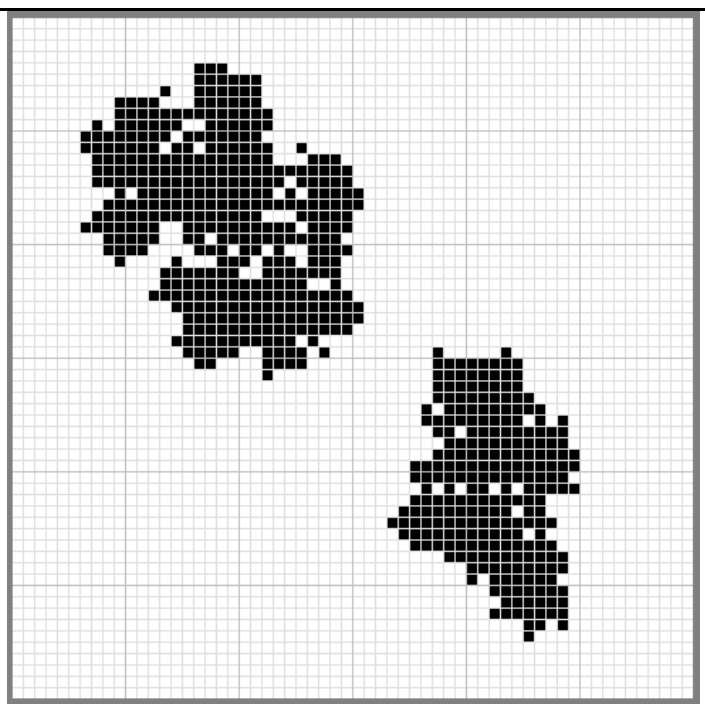<br>Adult 1, B3/S45678<br>*prob_mutation* = 0.0<br>*prob_selection* = 0.6<br>fitness = 616<br>black on target = 629<br>black off target = 13 | 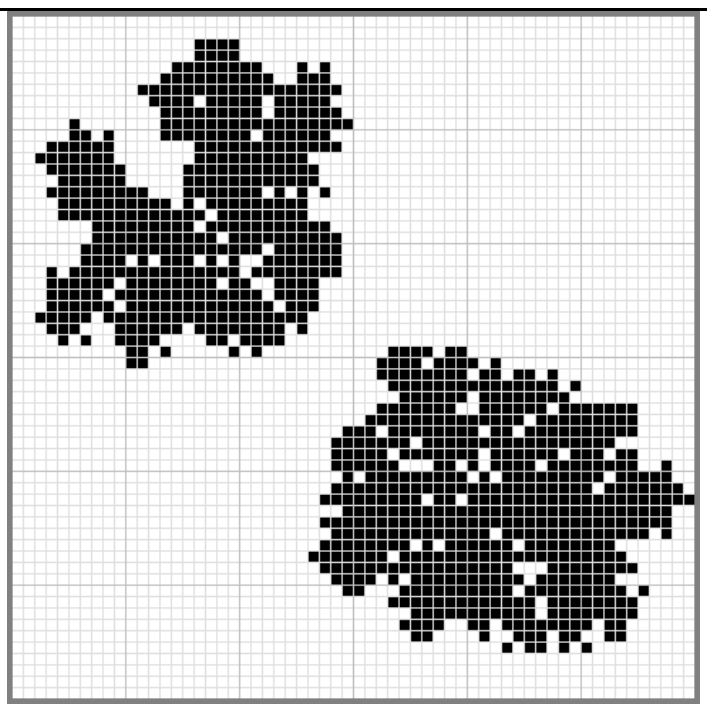<br>Adult 1, B3/S45678<br>*prob_mutation* = 0.1<br>*prob_selection* = 0.0<br>fitness = 951<br>black on target = 991<br>black off target = 40 |

Figure 7: Here we test whether we need both mutation and selection for good fitness scores.

In Figure 8, again we examine the impact of setting *prob_mutation* or *prob_selection* to zero, using Target 4 as a different test subject.

| | | |
|---|---|---|
| 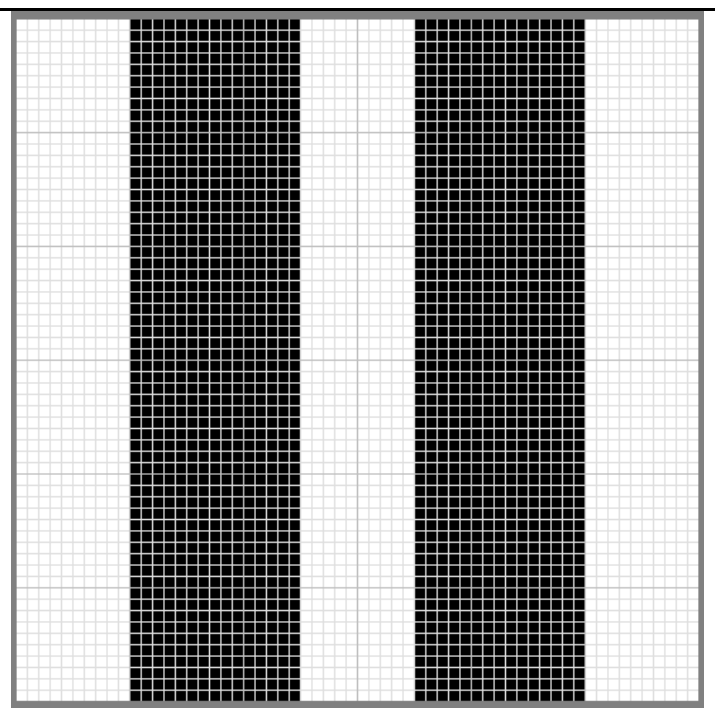<br>Target 4, 60 × 60 | 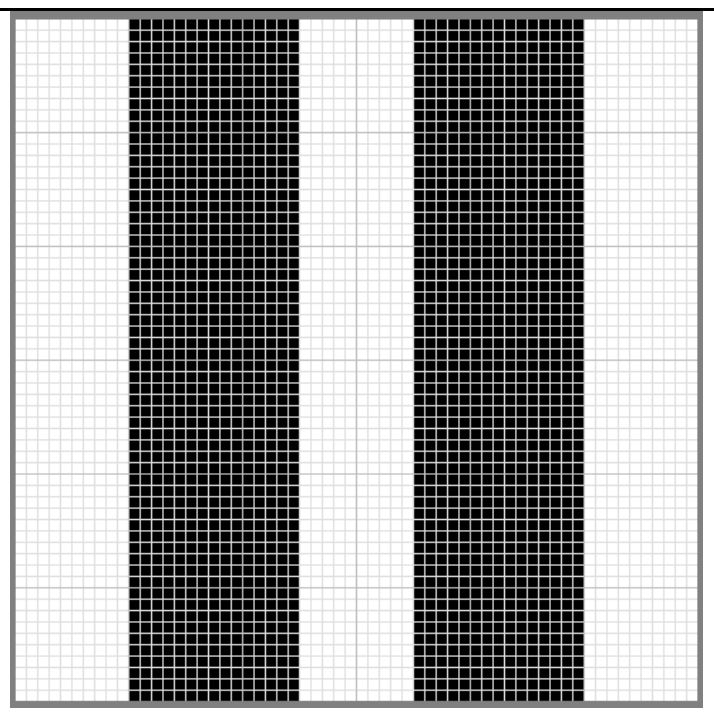<br>Target 4, 60 × 60 | 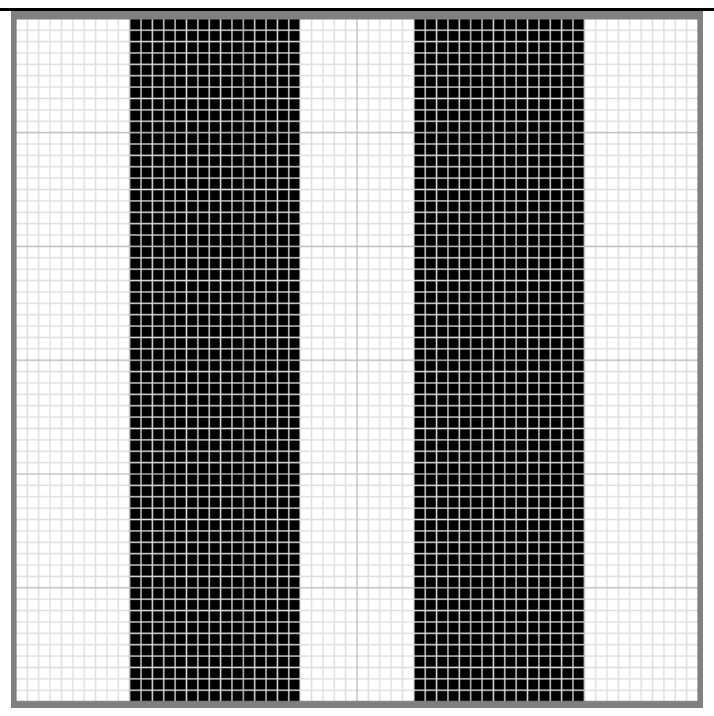<br>Target 4, 60 × 60 |
| 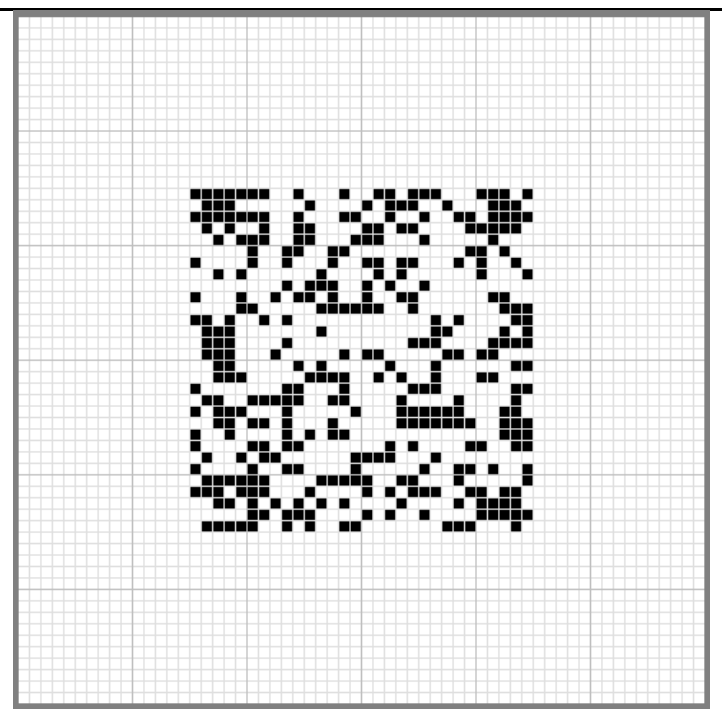<br>Seed 4, B3/S45678 | 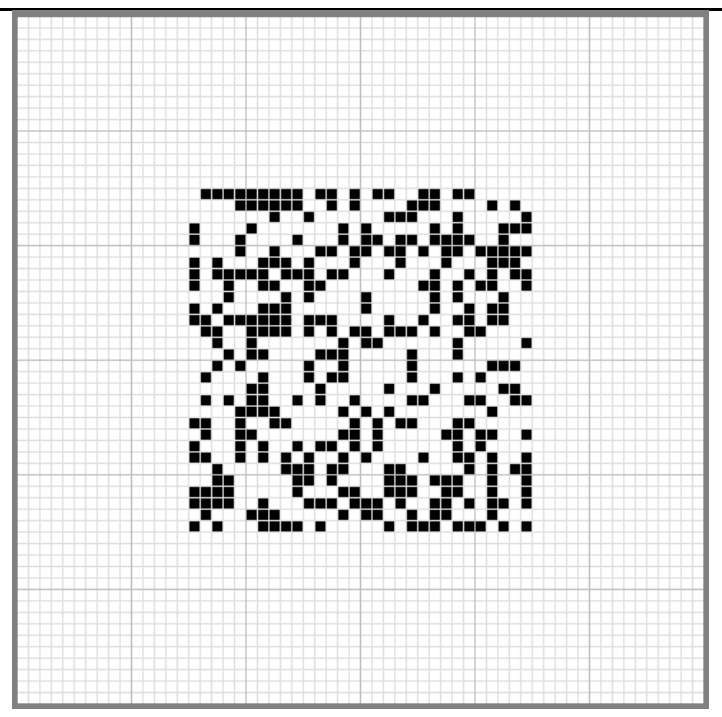<br>Seed 4, B3/S45678 | 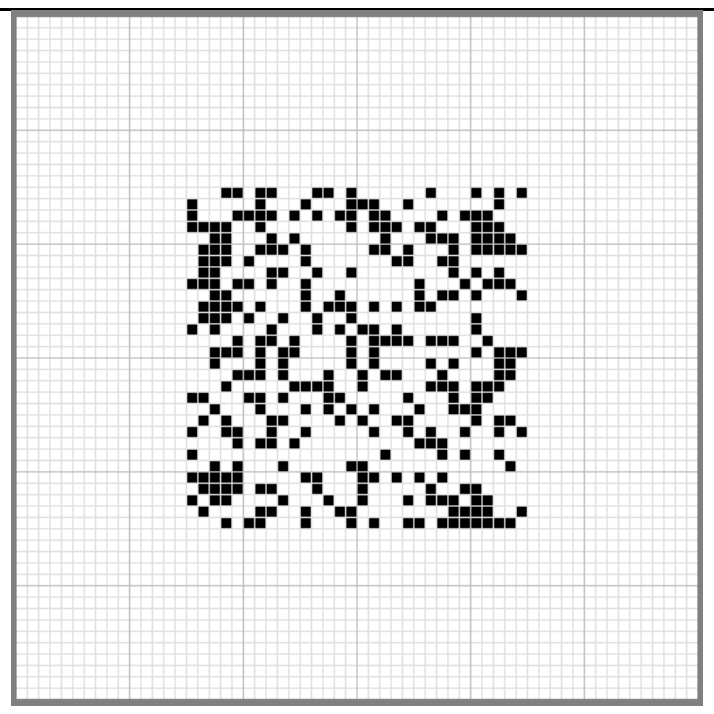<br>Seed 4, B3/S45678 |
| 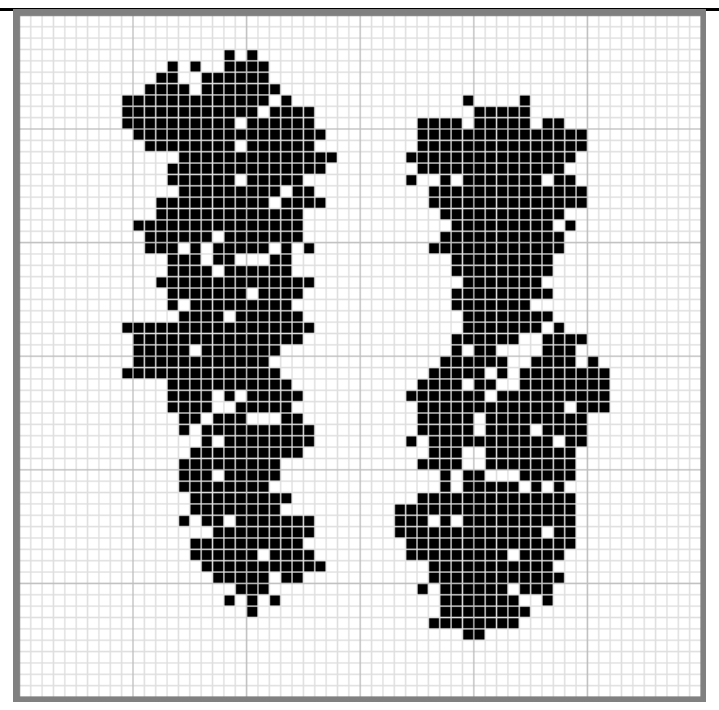<br>Adult 4, B3/S45678<br>*prob_mutation* = 0.1<br>*prob_selection* = 0.6<br>fitness = 957<br>black on target = 1006<br>black off target = 49 | 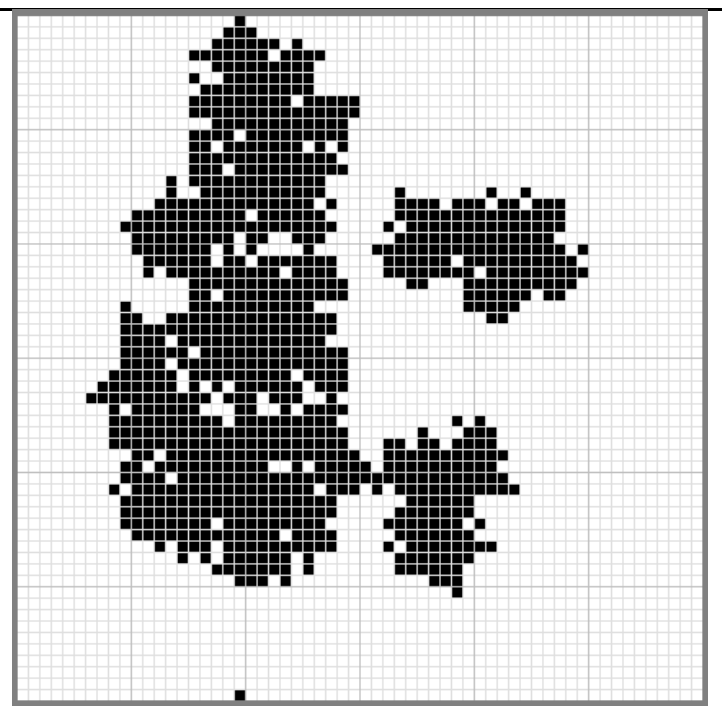<br>Adult 4, B3/S45678<br>*prob_mutation* = 0.0<br>*prob_selection* = 0.6<br>fitness = 538<br>black on target = 752<br>black off target = 214 | 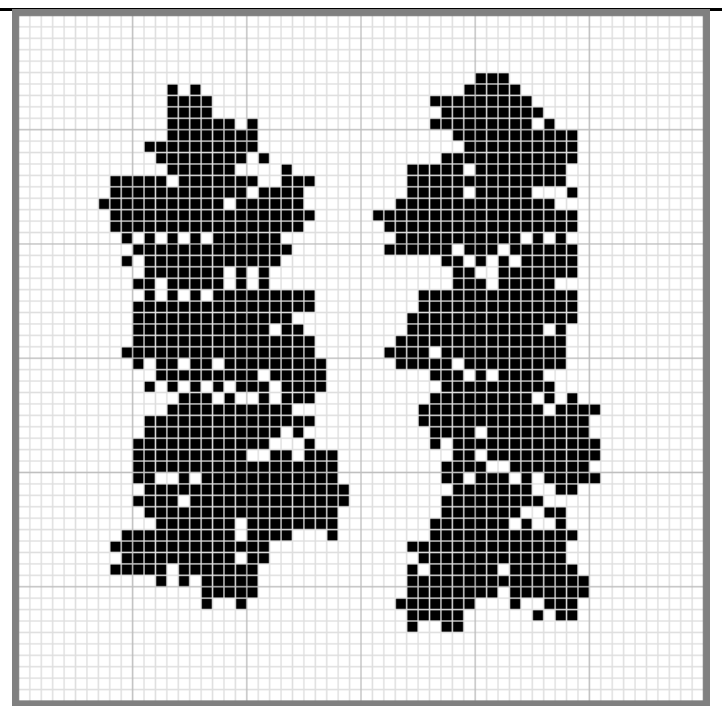<br>Adult 4, B3/S45678<br>*prob_mutation* = 0.1<br>*prob_selection* = 0.0<br>fitness = 938<br>black on target = 1035<br>black off target = 97 |

Figure 8: Again, we test whether we need both mutation and selection for good fitness scores.

Figures 7 and 8 show that mutation and selection working together perform better than mutation alone or selection alone. Mutation adds new patterns to the population, some of which may be useful and some which might not be useful. Selection removes seeds that perform poorly and adds seeds that perform well. Mutation and selection combined tend to perform better than mutation alone or selection alone.

# 5. Analysis

Perhaps the most interesting question that arises from this work is the nature of the relation between engineering and evolution. At first, it seems that engineering and evolving are almost opposites. Engineering requires prolonged, careful thought and design. Evolution seems random, thoughtless, and often wasteful. However, a closer look suggests that engineering often involves trying many different approaches and learning from many mistakes, until a solution is discovered. It can also be argued that evolution is sometimes very economical, for example, when it preserves pieces of DNA that are not currently useful, but may be useful in the future (Long, VanKuren, Chen, and Vibranovski, 2013).

Cellular automata are a kind of bridge between engineering and evolution. Cellular automata were developed in mathematics, but they quickly moved away from the Go board and into early computers. On the other hand, cellular automata are not like most computers. Cellular automata prefer two dimensions in a grid format, whereas most computers prefer a one-dimensional stream of alphabetical and numerical characters.

The Game of Life began in 1970. Many of the early patterns in the Game of Life were discovered by creating a random mess of white and black squares and then allowing the patterns to move, grow, shrink, and change (Johnston and Greene, 2022). Interesting patterns were saved and used as building blocks for larger structures. This process seems much like the evolution of biological life.

Now, 54 years after the beginning of Life, working with cellular automata has become more of an engineering task (Johnston and Greene, 2022). Recent constructions in Life are quite complex and they often involve teams working together.

The work presented in this article is moving away from the current highly engineered cellular automata (Johnston and Greene, 2022), towards a more evolutionary approach, but the evolution is now automated, rather than manual (Simon, 2013; Spears, 1998). The black targets we use to guide evolution (see Section 4 above) are somewhat inelegant, but we expect that the future will bring some more refined approaches.

# 6. Related Work

In past work (Turney, 2020), we simulated Game of Life organisms that could gradually increase their fitness over many generations, by adding layers of different types of reproduction. The first layer was simple asexual reproduction. The second layer added a more complex form of asexual reproduction. The third layer added sexual reproduction. The fourth layer added symbiosis. The fitness of the organisms was defined simply as how large they grew within a fixed time limit (a fixed number of steps in the game).

In the current article, the reproduction of Life organisms is based on simple asexual reproduction. It is different from the earlier work (Turney, 2020) in that the evolution of the organism is guided by various *targets*. A target is used to calculate the fitness of an organism. Organisms that match well with the target are more likely to reproduce (with mutations) and organisms that do not match well tend to die (they leave the population). This is unlike the past work (Turney, 2020), where growth was the only goal.

Perhaps the most similar work to ours is *Evolving Interesting Initial Conditions for Cellular Automata of the Game of Life Type* (Alfonseca and Soler Gil, 2012). We are both exploring the Game of Life and we are both particularly interested in the initial conditions. The initial conditions determine the future results. In our case, the evolutionary algorithm generates an evolved seed by randomly creating many seeds, using mutation and selection, until it finds the seed that best approximates the target. In the case of Alfonseca and Soler Gil, they explore time-dependent rules and alternating rules. Both of us use a 60×60 toroid.

Alfonseca and Soler Gil wrote that "… each complete execution of the genetic algorithm takes over half an hour, because the CA used by the algorithm must be run for each set of initial conditions in the population through generations 1 to 54 to compute their fitness…" (see their page 60). This is somewhat similar to our approach. The main difference is that our *targets* are bands of black and white, whereas their interest is in common objects in the Game of Life, such as gliders, R-pentominoes, and exploders.

# 7. Future Work

In our experiments above, each seed grows for 100 steps and then becomes an adult. If we allowed the growth to continue past 100 steps, without any interference, the adult would gradually change. It would no longer conform to the shape of the target. Its fitness would gradually decrease. This could be viewed as the adult entering old age and eventually death.

It would be possible to allow the evolution of the adult to continue indefinitely, if the adult were constantly mutated, repairing any damage that occurs. Ongoing mutation and selection could maintain the adult form indefinitely. The adult would still change over time, but the target would continuously guide it back, so that the adult never wandered far from the target.

# 8. Conclusions

Johnston and Greene (2022) provide an excellent survey of research in the Game of Life. They describe early discoveries in Life, based on *random fumbling* (their term, not ours), followed by classifying structures into various basic types, then combining parts to make complex interactive wholes. Their book spans twelve chapters, more than four hundred pages of discoveries about Life. Since Life is known to be a universal computer, it seems that there should be no limit to the growth of our knowledge about Life. This is supported by the progress we have seen over more than 50 years of Life (Gardner, 1970).

Johnston and Greene (2022) describe various software tools that have been constructed to support constructions in Life, but evolutionary algorithms are not discussed. The word *evolution* occurs 58 times in their book, but only in the generic sense of *change*, not in the sense of *mutation and selection* (Spears, 1998; Simon, 2013). The field of Life has evolved by a kind of *manual* mutation and selection of Life structures, but there is little work on *computational* mutation and selection of Life structures. One explanation for this is that the Game of Life community is essentially an *engineering* community, interested in building things by step-by-step design, whereas *evolutionary* algorithms build structures by mutation and selection, which is more about biology than about engineering.

## Acknowledgments

Thanks to Andrew Trevorrow, Tom Rokicki, Tim Hutton, Dave Greene, Jason Summers, Maks Verver, Robert Munafo, Brenton Bostick, and Chris Rowett, for developing the Golly cellular automata software. Thanks to Saif Mohammad, David Nadeau, and Nate Gaylinn, for helpful comments and suggestions. Thanks to the two *Artificial Life* reviewers for their very helpful comments and advice.